\documentclass[3p]{elsarticle} 
\makeatletter
\def\ps@pprintTitle{%
 \let\@oddhead\@empty
 \let\@evenhead\@empty
 \def\@oddfoot{\centerline{\thepage}}%
 \let\@evenfoot\@oddfoot}
\makeatother

\date{}
\def\texpsfig#1#2#3{\vbox{\kern #3\hbox{\includegraphics{#1}\kern #2}}\typeout{(#1)}}

\usepackage{url,hyperref}
\usepackage{bbm}
\usepackage{eurosym}                           
\usepackage[latin1]{inputenc}                  
\usepackage{url}                               
\usepackage{longtable}                         
\usepackage{array}                             
\usepackage{graphicx,color}
\usepackage{amsthm}
\usepackage{amsbsy}
\usepackage{subfig}
\usepackage[export]{adjustbox}
\usepackage{float}
\usepackage{tikz}
\usetikzlibrary{matrix,calc}
\usepackage{epsfig}

\usepackage{amssymb}
\usepackage{amsmath}
\usepackage{enumerate}
\usepackage{graphicx,color}
\usepackage{epsfig}
\usepackage{svg}
\usepackage[ruled,vlined]{algorithm2e}
\usepackage{multirow,bigdelim}
\usepackage{natbib}
\usepackage{array}
\usepackage{booktabs}

\usepackage[normalem]{ulem} 

\theoremstyle{plain}
\newtheorem{thm}{Theorem}[section]

\newtheorem{dfn}[thm]{Definition}

\newtheorem{rem}[thm]{Remark}
\theoremstyle{remark}

\theoremstyle{plain}

\theoremstyle{definition}

\newcommand{\w}{{\rm w}}
\newcommand{\W}{{\rm W}}
\def\b{\rm b}
\def\u{\rm u}
\def\w{\rm w}
\def\W{\rm W}
\def\R{\mathbb{ R}}             
\def\E{\mathbb{ E}}             

\def\P{\mathbb{ P}}             

\def\N{\mathcal{N}}             

\renewcommand{\d}{{\rm d}}      
\def\x{\rm x}
\def\dt{{\rm d}t}

\def\tttheta{{\theta}}
\def\ttheta{\theta}
\def\Dt{\Delta t}

\def\1{{\mathbbm{1}}}            

\theoremstyle{plain}

\usepackage[margin=1cm]{caption}
\newcommand{\zbox}[1]{
\noindent
\begin{center}
\resizebox{0.95\textwidth}{!}{
\framebox[14.5cm]{
\begin{minipage}{14cm}
#1
\end{minipage}
}
}
\end{center}
}

\numberwithin{equation}{section}	     

\title{Fast Sampling from Time-Integrated Bridges using Deep Learning}

\begin{document}

\author[1]{Leonardo Perotti\corref{cor1}}
\ead{Leonardo.Perotti@mail.polimi.it}
\author[2,3]{Lech A.~Grzelak}
\ead{L.A.Grzelak@uu.nl}
\cortext[cor1]{Corresponding author.}
\address[1]{Politecnico of Milan, Milan, Italy}
\address[2]{Utrecht University, Mathematical Institute, Utrecht, the Netherlands}
\address[3]{Rabobank, Utrecht, the Netherlands}

\begin{abstract}
    \noindent We propose a methodology to sample from time-integrated stochastic bridges, namely random variables defined as $\int_{t_1}^{t_2} f(Y(t))\d t$ conditioned on $Y(t_1)\!=\!a$ and $Y(t_2)\!=\!b$, with $a,b\in\R$.
    The techniques developed in \cite{GrWiSuOo} -- the Stochastic Collocation Monte Carlo sampler -- and in \cite{LiGrOo} -- the Seven-League scheme -- are applied for this purpose. Notably, the distribution of the time-integrated bridge is approximated utilizing a polynomial chaos expansion built on a suitable set of stochastic collocation points. Furthermore, artificial neural networks are employed to learn the collocation points. The result is a robust, data-driven procedure for the Monte Carlo sampling from conditional time-integrated processes, which guarantees high accuracy and generates thousands of samples in milliseconds. Applications, with a focus on finance, are presented here as well.
\end{abstract}

\begin{keyword}
  Stochastic Collocation Monte Carlo\sep
  Stochastic Bridge\sep
  Integrated Log-normal\sep Sampling from Heston/from SABR model \sep Artificial Neural Network\sep Seven-League Scheme.
\end{keyword}
\maketitle

\section{Introduction}  
\label{sec: intro}

This article focuses on \emph{stochastic bridges}, namely processes whose initial and final values are known a priori. Bridges have several applications, e.g., they are used in the study of \emph{time series} as stochastic interpolation rules; namely, they are employed to generate data with a higher sampling rate starting from a set with a lower sampling rate \cite{ScCh}.
In particular, in this article, we consider a common transformation of a stochastic bridge, it's integral over time. We call this class of random variables \emph{time-integrated bridges} or \emph{conditional time-integrated processes}. 

Given a generic stochastic process, it is always possible to define the corresponding bridge between two boundary conditions. 
However, sampling from stochastic bridges is an involved operation. Except for the special cases, the closed or semi-analytical form for the cumulative distribution function (CDF) of conditional distributions does not exist. Thus, on the one hand, it is not possible to perform exact sampling from the desired distribution (because of the lack of the CDF); on the other, the classical Monte Carlo (MC) schemes, as Euler-Maruyama or Milstein discretizations, are not applicable (the boundary conditions cannot be easily imposed). The proposed approach constitutes a generic framework for sampling from conditional distributions, stochastic bridges and their transformations.

Particularly, we propose a fast and accurate alternative to the classical MC sampler for the above-mentioned class of {conditional time-integrated processes} (or {time-integrated bridges}), i.e.
\begin{equation*}
    \int_{t_1}^{t_2} f(Y(t))\d t\:\Big|\, Y(t_1)=a, Y(t_2)=b,\quad a,b\in\R,\quad f\in\mathcal{C}^0(\mathcal{D}),\:\: \mathcal{D}\subset \mathbb{R},
\end{equation*}
where $Y(t)$ is a stochastic process defined between the times $t_1$ and $t_2$.
Observe that from a numerical viewpoint, the integral over time of a process is handled as a summation over a time grid. For instance, the so-called trapezoidal rule for the integral of a function $f$ is given by the approximation
\begin{equation*}
    \int_{t_1}^{t_2} f(\tau) \d \tau \approx \frac{\Delta \tau}{2}\big(f(\tau_0) + f(\tau_N)\big) + \Delta \tau \sum_{i=1}^{N-1} f(\tau_i),
\end{equation*}
with $t_1=\tau_0<\tau_1<\dots<\tau_N=t_2$, $\Delta \tau=\tau_1 - \tau_0 = \tau_i - \tau_{i-1}$ for each $i$. Hence, the same idea developed here can be employed when dealing with a finite sum of random variables parameterized in time.

The literature about MC sampling from conditional distributions is not very rich. Notably, no general sampling scheme is available for stochastic bridges and their transformations. Our work aims to develop and formalize a procedure that allows performing sampling from time-integrated stochastic bridges.

The method relies on two main ideas. First, the time-integrated bridge distribution is ``compressed'' into a few collocation points (CPs) through the \emph{Stochastic Collocation Monte Carlo} (SCMC) technique \cite{GrWiSuOo}; then, artificial neural networks (ANNs) are employed to ``learn'' the CPs allowing for quick recovery \cite{LiGrOo}. Compression and deep learning techniques allow developing a numerical scheme to sample from any choice of the time-integrated bridge in a fast and accurate fashion.
The SCMC model allows fast simulation from most of the common distributions, and it can be seen as a powerful and elegant alternative to the classical MC simulation when dealing with expensive distributions, whose CDF is available (at least some approximation). Moreover, the application of this method to the exact simulation in \cite{BrKa} allows fast and accurate simulation in the Heston framework. Here, the SCMC technique guarantees a compact way to store all the relevant information concerning MC paths (``compression'') and a procedure to recover them when needed (``decompression''). 
On the other hand, in the \emph{Seven League} (7L) scheme \cite{LiGrOo} the problem of ``large time step'' Monte Carlo simulation is tackled utilizing a polynomial chaos expansion method based on appropriate stochastic CPs. The CPs are obtained through a regression problem solved with \emph{deep learning} techniques as ANNs. In this work, we learn the CPs solving a regression problem with ANNs in the same flavour as in \cite{LiGrOo}. For this reason, we may refer to the proposed methodology as \emph{Seven-League Scheme} (7L) if there is no ambiguity with the original one.
For the sake of completeness, we observe that {deep learning} techniques have been employed more and more in different frameworks and, in particular, to solve differential equations. For instance, Physics-Informed neural networks \cite{RaPeKa} are applied to the problem of solving ordinary and partial differential equations \cite{BeChJe,HaJeE}, obtaining satisfactory results in terms of computational speed and allowing to handle high-dimensional problems unsolvable easily with the classical numerical methods, whereas in \cite{LiGrOo}, ANNs are employed to solve SDEs.
Hence, in the Monte Carlo sampling framework, ANNs are novel tools that facilitate a speedup of long computations while maintaining high accuracy in the results. Therefore, the approach proposed provides an original example of the application of deep learning in the framework of Monte Carlo sampling, and it opens to further research directions.

The remainder of the paper is organized as follows. Section \ref{sec: model framework} introduces the model framework, namely the SDE setting and the notation. Particularly, in Section \ref{ssec: financial motivations}, an insight of financial motivations is provided, while in Section \ref{ssec: general framework}, the general framework is presented. Then, details about the collocation technique are given in Section \ref{ssec: SCMC}, with particular attention to the case of time-integrated bridges (Section \ref{sssec: SCMC for Z}). In Section \ref{sec: methodology} the methodology is presented. First the ``off-line'' stage (Section \ref{ssec: off-line}) is described. In this section all the details of the so-called ``compression'' part (Section \ref{sssec: compr}) are given, together with an accurate explanation of the deep learning techniques employed (Section \ref{sssec: ANN}). The ``on-line'' stage follows in Section \ref{ssec: online}. The sampling from time-integrated bridges uses the ANN trained off-line (Section \ref{sssec: CP computation}) coupled with an appropriate ``decompression'' technique (Section \ref{sssec: decompr}). A brief discussion on the errors is given in Section \ref{ssec: error}. The methodology is tested, in Section \ref{sec: appl}, on different models (Section \ref{ssec: models}), such as the conditional time-integrated \emph{Arithmetic Brownian Motion} (ABM), the conditional time-integrated \emph{Geometric Brownian Motion} (GBM) and the conditional time-integrated \emph{Cox-Ingersoll-Ross} (CIR) model. Financial applications are shown in Section \ref{ssec: fin applic}. Eventually, Section \ref{sec: conclusion} concludes.

\section{Motivations from finance and modelling framework}
\label{sec: model framework}

Before to introduce the general model framework in which our work will be developed, let us recall that the goal of this article is to propose a scheme -- alternative to plain Monte Carlo simulation -- which allows to sample from any kind of time-integrated stochastic bridge, or more in general any kind of deterministic transform of a stochastic bridge, in a fast and accurate way. As pointed out previously, such a methodology finds several real applications, as shown in the following section.

\subsection{Financial motivations}
\label{ssec: financial motivations}
Conditional time-integrated processes appear in different fields, particularly in quantitative finance. Two natural financial applications of the method concern well-known stochastic volatility (SV) models as the Heston model and the Stochastic Alpha Beta Rho (SABR) model. Indeed, in order to sample from these models it is necessary to deal with conditional time-integrated processes. An insight of how to employ the methodology proposed is provided in this work in Section \ref{ssec: fin applic}.

\subsubsection{Heston Stochastic Volatility model}
\label{sssec: intro Heston}
The Heston model -- used to describe the evolution of a stock price in time -- is characterized by \emph{stochastic variance}, namely the variance is not anymore a constant value (as it was in the Black-Scholes model), and it is not even a deterministic function of the time, but it is a stochastic process itself.

Under the Heston model assumptions, the sampling of the log-price of stock requires the computation of a conditional time-integrated process \cite{An, BrKa}. In \cite{BrKa} is proposed an \emph{exact simulation} for the log-price process, which requires the computation of the time-integrated variance conditioned to its initial and final values.

Particularly, the dynamics of the variance process $v(t)$ between the times $t_1$ and $t_2$, $0 \leq t_1 < t_2$, is expressed in terms of a Cox-Ingersoll-Ross (CIR) model by the following SDE
\begin{equation}
\label{eq: variance Heston}
    \d v(t) = \kappa (\overline{v}-v(t)) \dt + \gamma \sqrt{v(t)} \d W_v(t),
\end{equation}
where the (constant) parameters $\kappa$, $\overline{v}$, $\gamma$ represent respectively the \emph{mean-reverting speed}, the \emph{long-term variance} and the \emph{volatility of volatility (vol-of-vol)}, while $W_v(t)$ is a standard Brownian motion.

As a generalization of the Black-Scholes model, defining $X(t)$ as the log-price of the stock, the Heston model dynamics (with constant parameters) reads
\begin{equation}
\label{eq: Heston model dep}
\begin{aligned}
    &\d X(t)=\big(r-\frac{1}{2}v(t)\big)\d t + \sqrt{v(t)}\d W_X(t),
\end{aligned}
\end{equation}
with $v(t)$ as in Equation (\ref{eq: variance Heston}),  $W_X(t)$ and $W_v(t)$ correlated Brownian motions with correlation coefficient $\rho$, i.e. $\d W_X(t)\d W_v(t)=\rho\d t$.
The system of SDEs defined by Equations (\ref{eq: variance Heston}) and (\ref{eq: Heston model dep}) can be rewritten in terms of independent Brownian motions $\widetilde{W}_X(t)$ and $\widetilde{W}_v(t)$ as
\begin{equation}
\label{eq: Heston model indep}
\begin{aligned}
    & \d X(t)=\big(r-\frac{1}{2}v(t)\big)\d t + \rho\sqrt{v(t)}\d \widetilde{W}_v(t)+\sqrt{1-\rho^2}\sqrt{v(t)}\d \widetilde{W}_X(t),\\
    & \d v(t) = \kappa (\overline{v}-v(t)) \dt + \gamma \sqrt{v(t)} \d \widetilde{W}_v(t).
\end{aligned}
\end{equation}
Exploiting the integral form, by substitution, we obtain
\begin{equation}
\label{eq: Heston log-price discretization final}
\begin{aligned}
    X(t_2) & = X(t_1) + \Big(r-\frac{\rho\kappa\overline{v}}{\gamma}\Big)(t_2-t_1) + \sqrt{1-\rho^2} \int_{t_1}^{t_2} \sqrt{v(t)}\d\widetilde{W}_X(t)\\
    &+\Big(\frac{\rho\kappa}{\gamma}-\frac{1}{2}\Big)\boxed{\int_{t_1}^{t_2} v(t) \d t}+ \frac{\rho}{\gamma}\big( \boxed{v(t_2)} - \boxed{v(t_1)}\big).
\end{aligned}
\end{equation}
It is enough to look at the last three terms to observe that the problem of sampling from a time-integrated process conditioned to its initial and final realization arises naturally from this framework. Indeed, in order to compute $X(t_2)|X(t_1)$ we need to sample together $v(t_1)$, $v(t_2)$ and $\int_{t_1}^{t_2} v(t) \d t$.

Eventually, it can be shown that the stochastic integral $\int_{t_1}^{t_2} \sqrt{v(t)}\d\widetilde{W}_X(t)$ conditional to the path of the process $v(t)$ is distributed as normal random variable with mean 0 and variance $\int_{t_1}^{t_2} v(t) \d t$ \cite{An}. Therefore, being able to sample from conditional time-integrated variance process is sufficient to perform highly accurate numerical simulations from the Heston model.
\begin{rem}[Exact sampling from Heston model]
In \cite{BrKa} is proposed a procedure to perform exact sampling under the Heston model hypothesis. The distribution of the conditional time-integrated variance process is obtained employing a computationally expensive Fourier series inversion. Even though the procedure provided here is obtained starting from Monte Carlo simulations, such a technique could be naturally employed and extended to improve the exact sampling given in \cite{BrKa}, speeding-up significantly the overall computation.
\end{rem}

\subsubsection{Stochastic Alpha Beta Rho (SABR) model}
\label{sssec: intro SABR}
The SABR model is employed to describe the dynamics of a stock forward price $S(t)$, $0\leq t_1\leq t \leq t_2$. Different from the Heston framework, under these model assumptions, no exact simulation is available. Nonetheless, it is possible to produce an accurate approximation. The method requires an ``expensive'' Fourier-based inversion of (an approximation of) the conditional price process cumulative distribution function (CDF) and involves copulas to handle the dependence relationships \cite{LeGrOo2, LeGrOo}.

For the general SABR model the dynamics can be expressed through the following system of SDEs
\begin{equation}
\begin{aligned}
    &\d S(t) = \sigma(t)S^\beta(t) \d W_S(t),\\
    &\d \sigma(t) = \alpha \sigma(t) \d W_\sigma(t),
    \end{aligned}
\end{equation}
where $\alpha>0$, $0\leq\beta\leq 1$ are model parameters and $W_S(t)$ and $W_\sigma(t)$ are correlated Brownian motions with correlation coefficient $\rho$, i.e. $\d W_S(t)\d W_\sigma(t)=\rho\d t$.

In this framework, it is possible to perform numerical simulation of the solution $S(t_2)|S(t_1)$ inverting the conditional CDF \cite{Is}
\begin{equation}
\label{eq: SABR conditional CDF}
    \P\Big[S(t_2)\leq z\big|S(t_1)>0, \sigma(t_1), \sigma(t_2),\int_{t_1}^{t_2}\sigma^2(t)\dt\Big]=1-\chi^2(\overline{a};\overline{b}, \overline{c}),
\end{equation}
where $\chi^2(x;\delta, \lambda)$ is the non-central chi-square CDF (with $\delta$ degrees of freedom and non-centrality parameter $\lambda$) and
\begin{equation}
    \begin{aligned}
    &\overline{a}=\frac{1}{\nu(t_1,t_2)}\bigg(\frac{S(t_1)^{1-\beta}}{1-\beta}+\frac{\rho}{\alpha}\big(\boxed{\sigma(t_2)}-\boxed{\sigma(t_1)}\big)\bigg)^2,\quad
    &\overline{b} = 2-\frac{1-2\beta-\rho^2(1-\beta)}{(1-\beta)(1-\rho^2)},\\
    &\overline{c}=\frac{z^{2(1-\beta)}}{(1-\beta)^2\nu(t_1,t_2)},\quad
    &\nu(t_1,t_2)=(1-\rho^2)\boxed{\int_{t_1}^{t_2}\sigma^2(t)\dt}.
    \end{aligned}
\end{equation}
Inspecting $\overline{a}$ and $\nu(t_1,t_2)$, in the same fashion of the Heston framework case, there is need to sample from time-integrated (squared) volatility process conditioned on the initial and final values, $\sigma(t_1)$ and $\sigma(t_2)$. Although other methodologies for this sampling have been implemented \cite{LeGrOo2, LeGrOo}, the result of this work has an immediate application here and allows to achieve accurate simulation keeping low the computational effort.

\subsection{General framework}
\label{ssec: general framework}

Let us move, now, to the general framework. We begin fixing some notation.
Given $0 \leq t_1 < t_2$ and the set of model parameters $\ttheta_0\in\Theta\subset \R^N$, $N\geq 1$, consider the real process $Y(t)$ whose dynamics are driven by the following generic SDE
\begin{equation}
\label{eq: dynamics theta}
    \d Y(t) = \Bar{\mu}(t,Y(t);\ttheta_0) \d t + \Bar{\sigma}(t,Y(t);\ttheta_0) \d W(t),
\end{equation}
where $\Bar{\mu}(t,y;\ttheta_0)$ and $\Bar{\sigma}(t,y;\ttheta_0)$ are smooth deterministic real functions from $[t_1,t_2]\times \R$, respectively the \emph{drift} and the \emph{diffusion} terms of the SDE and $W(t)$ is a standard Brownian motion.
Once an initial condition is specified, Equation (\ref{eq: dynamics theta}) is equivalent to the integral equation
\begin{equation}
    \label{eq: dynamics integral}
    Y(t) = Y(t_1) + \int_{t_1}^t \Bar{\mu}\big(s,Y(s);{\ttheta_0}\big) \d s + \int_{t_1}^t \Bar{\sigma}\big(s,Y(s);{\ttheta_0}\big) \d W(s),
\end{equation}
where the latter term is the so-called \emph{It\^o's stochastic integral}. 

\begin{figure}[t]%
    \centering
    \subfloat[\centering]{\includegraphics[width=7.0cm]{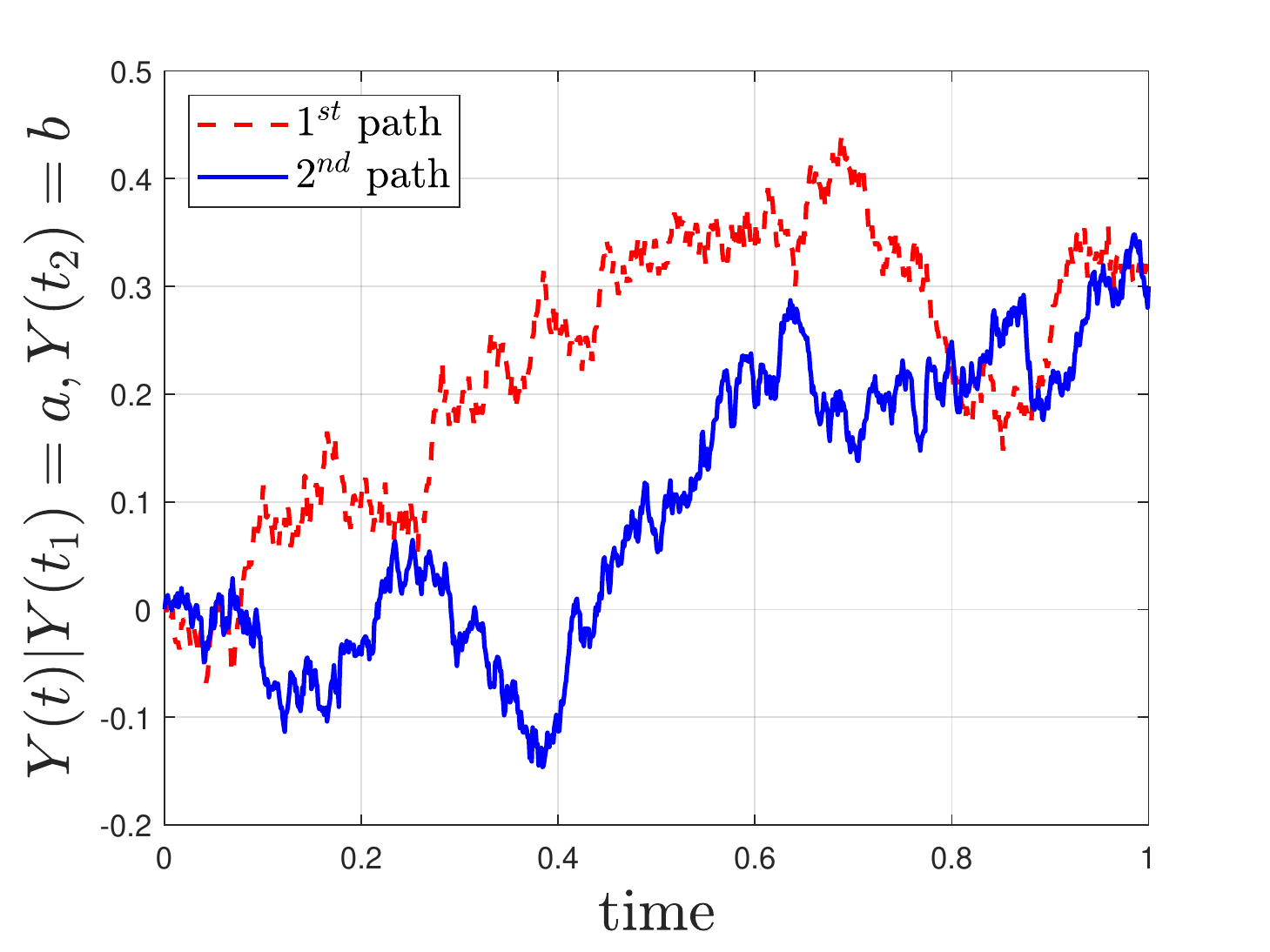} }%
    ~
    \subfloat[\centering]{{\includegraphics[width=7.0cm]{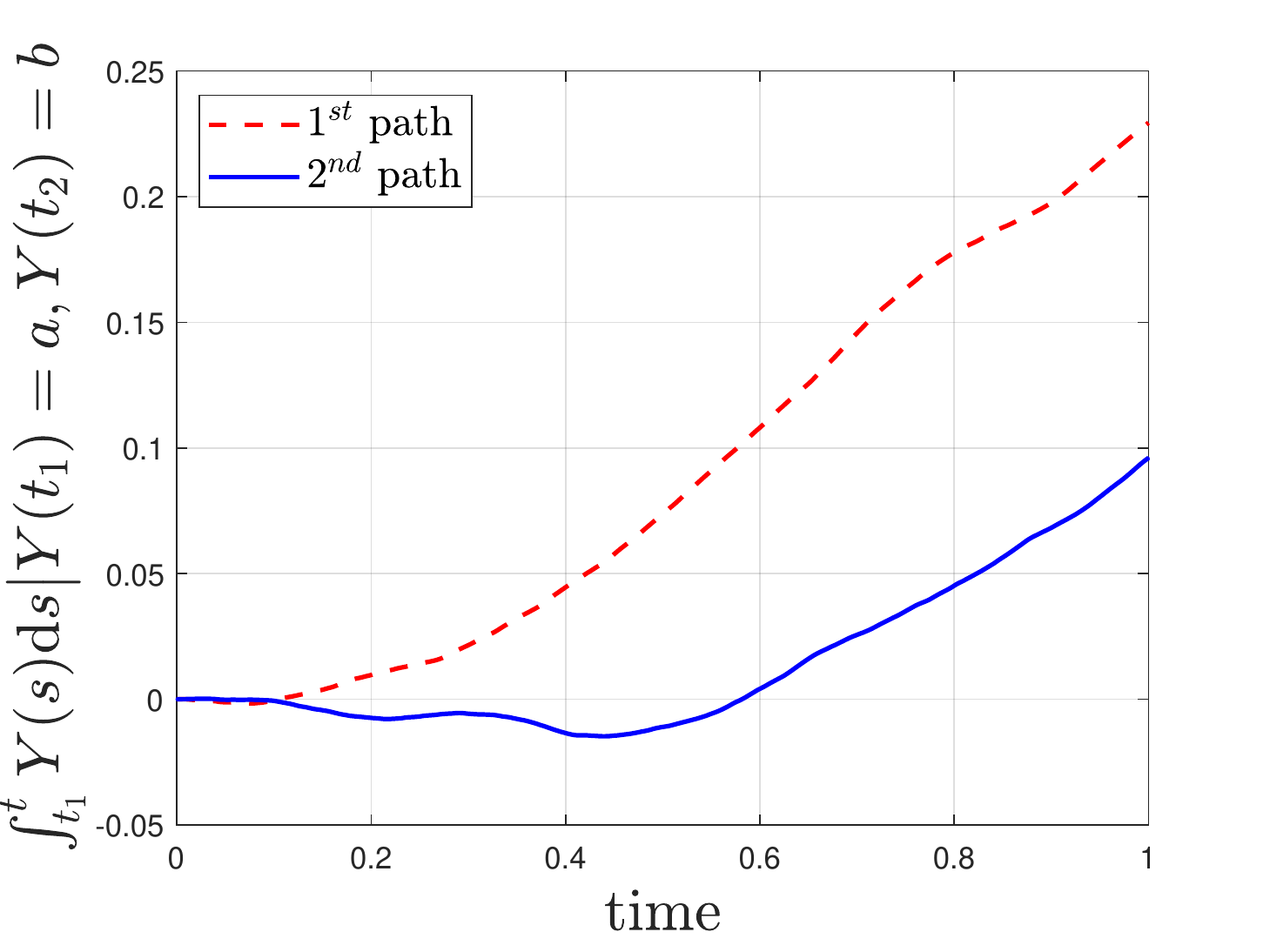} }}%
    \vspace{-0.35cm}
    \caption{\footnotesize Conditional ABM with the corresponding integral over time between $t_1=0$ and $t_2=1$, $\mu$ arbitrary chosen and $\sigma=0.3$. The boundary conditions are specified as $a=0.0$ and $b=0.3$.}
    \label{fig: Integrated ABB}%
    \vspace{-0.25cm}
\end{figure}

We observe that in general, given a time $t \in [t_1,t_2]$, an analytic expression for $Y(t)$ does not exist. In most of the cases it is not even possible to have an analytic form of the CDF of $Y(t)$. This makes the sampling from a generic process an involved operation, since the most common techniques are based on the inversion of the CDF.
Moreover, we are not interested in generic processes, but our focus is on stochastic bridges, namely stochastic processes conditioned on their initial and final values or boundary conditions, and in particular, the target distribution -- the one we want to sample from -- is the integral over time of a stochastic bridge (see Figure \ref{fig: Integrated ABB}). 
Let us formalize this concept in the following definition.
\begin{dfn}[Conditional time-integrated process]
Let us consider $0 \leq t_1 < t_2$ and the process $Y(t)$ be given in terms of the dynamics in Equation (\ref{eq: dynamics theta}), with model parameters $\theta_0$. Moreover, let us define $\theta:=\theta_0\cup \{t_1,t_2\}$, with $t_2:= t_2 - t_1$.
For $a,b\in\R$ and $f$, a deterministic function, we call \emph{conditional time-integrated (transformed) process} -- or \emph{time-integrated (transformed) bridge} -- \emph{between $a$ (or $f(a)$) and $b$ (or $f(b)$)} the quantity
\begin{equation}
    Z(\theta|a, b):=\int_{t_1}^{t_2} f(Y(t)) \d t\:\bigg| \:Y(t_1)=a, Y(t_2)=b.
    \label{eq: integrated bridge definition}
\end{equation}
\end{dfn}

In light of this definition, we can reformulate the purpose of this work as develop a methodology to efficiently sample from the conditional time-integrated process $Z(\theta|a,b)$. In particular, we want to be able to handle a general case, which means that the method have to take as ``input'' any choice of parameters\footnote{Here we can see the boundary conditions $(a,b)$ just as two more parameters, to be added to $\theta$.} and return as ``output'' accurate MC samples from the target distribution $Z(\theta|a,b)$.

The methodology consists of two main parts.
As a first step, a one-time ``{off-line}'' stage, possibly computationally expensive, has to be implemented. Then, the result of the off-line phase is employed in the ``{on-line}'' phase. Mainly, during the on-line phase, the actual sampling from the time-integrated bridge is performed. Such a simulation is comparable to computationally expensive classical MC schemes in terms of accuracy, while the overall time needed is significantly reduced.
The most important part of the off-line stage is the ``training'' of a suitable function $H$, that is the calibration of a set of parameters which determines $H$ among the family of all the functions that can be obtained varying the parameters' set. In principle, such a function ``learns'' how to map quickly the tuple $\theta$ into a set of \emph{collocation points} stored in grids. These collocation points contain the information necessary to recover highly accurate samples from the distribution of $Z(\theta|a,b)$, for any pair of boundary conditions $(a,b)$. The ``compression'' of the whole information to perform MC simulation into the collocation points is based on the SCMC sampler \cite{GrWiSuOo}. Better insight into the SCMC technique is given in the next section. The learning phase, on the other hand, employs deep learning regression techniques based on artificial neural networks (ANNs), in the same flavour as in \cite{LiGrOo}.
As pointed out previously, the on-line phase is the actual sampling algorithm. Given the model parameters, the corresponding collocation points are computed employing the map $H$, trained during the off-line stage. Eventually, the MC samples are obtained from the collocation points through the ``decompression'' phase (the information collected in the grids is ``decompressed'' to generate back MC samples). For any choice of boundary conditions $(a,b)$, only the relevant information is extracted from the grids, and it is used for the sampling.

The interesting aspect is that the only computationally expensive operation, i.e. the training of the map $H$, is performed just once (and off-line!), whereas each time it is necessary to sample from $Z(\theta|a,b)$ (for any choice of $\ttheta$, $a$, $b$), only the on-line stage, which is extremely fast, is required. Moreover, an important fact, particularly for the implementation, is stated in the following remark.

\begin{rem}[Time-invariant Markovian processes]
\label{rem: time-inv Markov processes}
The general model framework presented so far is not common in applications. Indeed, usually, the processes we are interested in have two properties that allow simplifying the methodology. In particular, we almost always deal with \emph{time-invariant} and \emph{Markovian} stochastic processes. Roughly, a process is called ``Markovian'' if the information about the future available at present includes the one available from the past. On the other hand, a ``time-invariant'' process is such that, given $0\leq t_1 < t_2$, the distribution of the process at time $t_2$ does not depend on the time state $t_2$ itself, but only on the time variation $\Dt:=t_2-t_1$.
For this -- extremely common -- class of processes, there is no need to consider the pair of parameters $(t_1,t_2)$, or equivalently $(t_1, \Dt)$, but it is enough to consider the latter, namely $\Dt$, assuming implicitly $t_1=0$. As a result, for this class of processes, we have the simplified set of parameters $\theta:=\theta_0 \cup \{\Dt\}$. This is useful to ease the implementation.
\end{rem}

\subsection{Stochastic Collocation Monte Carlo}
\label{ssec: SCMC}
Before we describe the proposed methodology with its off-line and on-line phases, we briefly present the Stochastic Collocation Monte Carlo (SCMC) technique (see \cite{GrWiSuOo} for details).
A common way to sample from a given distribution is utilizing the inversion of its CDF. Indeed, given a random variable $Z$ and calling $F_Z(z)=\P[Z\leq z]$ its CDF, it is possible to draw from its distribution just by evaluating the inverse\footnote{We assume $F_Z$ strictly increasing, hence invertible.} CDF $F_Z^{-1}$ in $\Bar{U}$, where $\Bar{U}$ is a sample from a real random variable $U$ uniformly distributed in the interval $[0,1]$, namely $U\sim \mathcal{U}\big([0,1]\big)$. We remark that the ``bar-notation'' is employed every time we deal with realizations from a given random variable to distinguish them from the random variable itself.

The idea underlying this technique is provided by the identity
\begin{equation}
\label{eq: identity unif}
    F_Z(Z)\overset{d}{=}U=F_U(U),
\end{equation}
where the last equality is trivial, since $F_U(u)=\P[U\leq u]=u$, $u\in[0,1]$.
Equation (\ref{eq: identity unif}) can be generalized to different random variables. Indeed, let $\xi$ be a generic random variable with CDF $F_\xi$, the following similar identity holds
\begin{equation}
\label{eq: identity implicit}
    F_Z(Z)\overset{d}{=}F_\xi(\xi).
\end{equation}
From Equation (\ref{eq: identity implicit}) we can express the variable $Z$ as a proper transformation $g$ of $\xi$ defined by the relationship
\begin{equation}
\label{eq: identity explicit}
    Z\overset{d}{=}g(\xi):=F_Z^{-1}(F_\xi(\xi)).
\end{equation}
Therefore, if $g$ is known, the sampling from $Z$ can be easily done evaluating $g$ in the realization $\Bar{\xi}$ of $\xi$, i.e.
\begin{equation}
    \label{eq: identity sampling}
    \Bar{Z}= g(\Bar{\xi}),
\end{equation}
is a realization from $Z$.
Observe that to have a benefit from Equation (\ref{eq: identity sampling}), the sample from the random variable $\xi$ must be easily obtainable; otherwise, we are only moving the problem and not solving it. As a matter of fact, (standard) normal distribution is tabulated in most computing tools, making it a suitable candidate for the ``cheap'' random variable $\xi$.

Nonetheless, in general, the map $g$ is not available, or it is computationally expensive to achieve. Indeed, as we can see in Equation (\ref{eq: identity explicit}), to obtain $g$, we have to invert the CDF of $Z$. This operation cannot always be performed analytically or simply, and it is too expensive from a computational viewpoint. This may be a problem if there is the need to repeat the inversion many times, as we can expect when a MC simulation is run for any purpose.

The SCMC technique provides a possible solution to this issue. Indeed, it allows recovering an accurate approximation of $g$ utilizing a suitable interpolating function. However, the choice of interpolation is non-trivial. Indeed, several options exist, with different features. Possible choices are Lagrange interpolation (easy to implement, but unfortunately, it may lack monotonicity), Chebyshev interpolation (based on trigonometric polynomials), and piece-wise cubic spline interpolation.

In the case of Lagrange interpolation (with an $M-1$ degree polynomial), the polynomial is built requiring only $M$ inversions of the expensive CDF $F_Z$ at the $M$ {collocation points} (CPs) \cite{GrWiSuOo}. This guarantees a reduced computational load since the number of expensive operations, i.e., the number of the CDF's inversions does not depend on the magnitude of the MC simulation (roughly, it does not depend on the number of samples generated).

In particular, the set $\{\xi_k\}_{k=1}^M$ of original CPs (for the ``cheap'' random variable $\xi$) is specified a priori, then the $M$ pairs $(\xi_k, z_k)$, $k=1,\dots,M$, are computed with second entry given by
\begin{equation}
\label{eq: computation CPs}
z_k=F_Z^{-1}(F_\xi(\xi_k)).
\end{equation}
We emphasize that this computation may be expensive, but it is performed only a small and constant number of times, that is $M$. 
Notice that in order to avoid ambiguity, we call the $\xi_k$ original CPs, whereas we address to the $z_k$ as (stochastic) CPs.

\begin{figure}[b!]%
    \centering
    \subfloat[\centering]{\includegraphics[width=7cm]{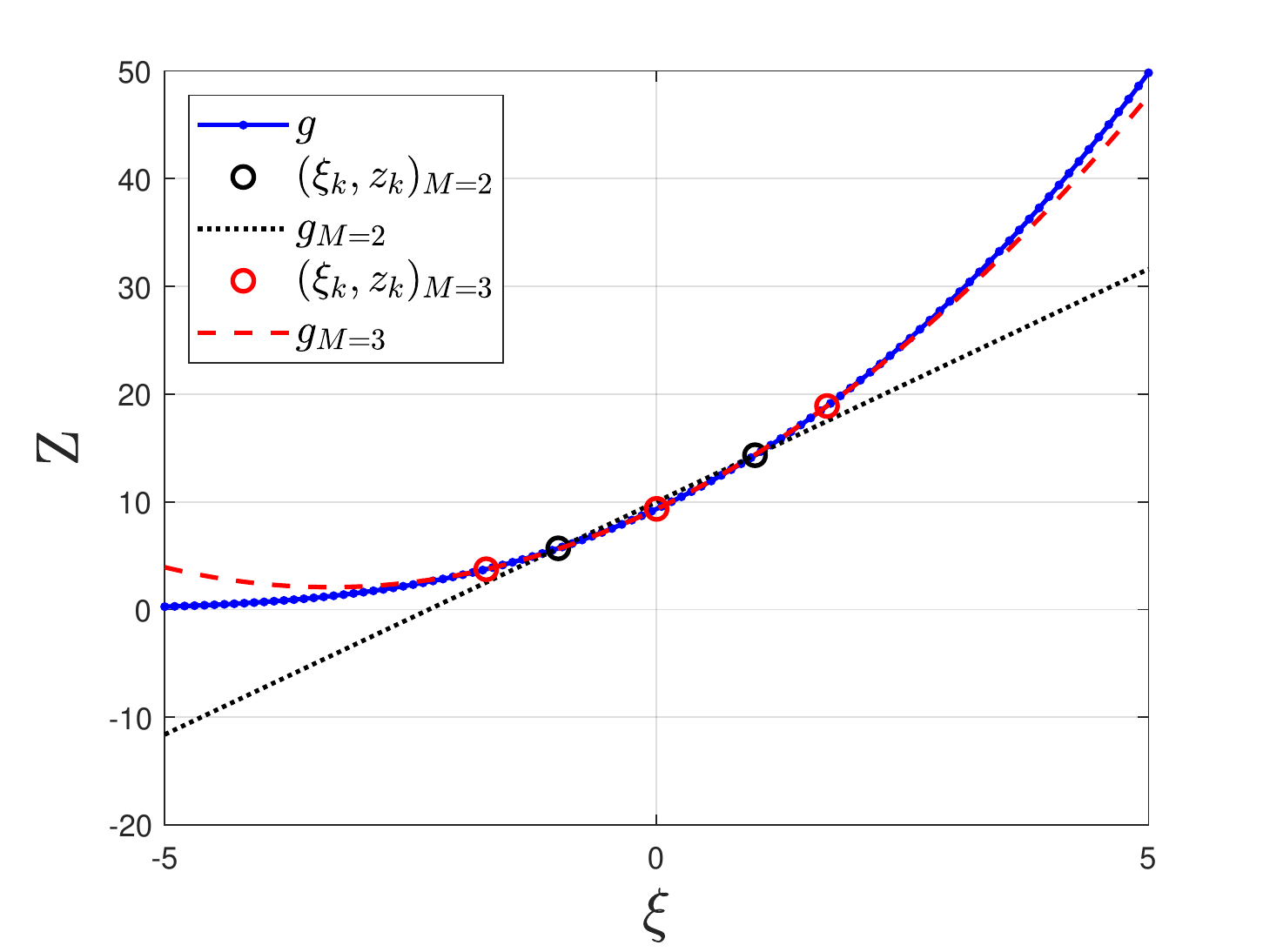} }%
    ~
    \subfloat[\centering]{{\includegraphics[width=7cm]{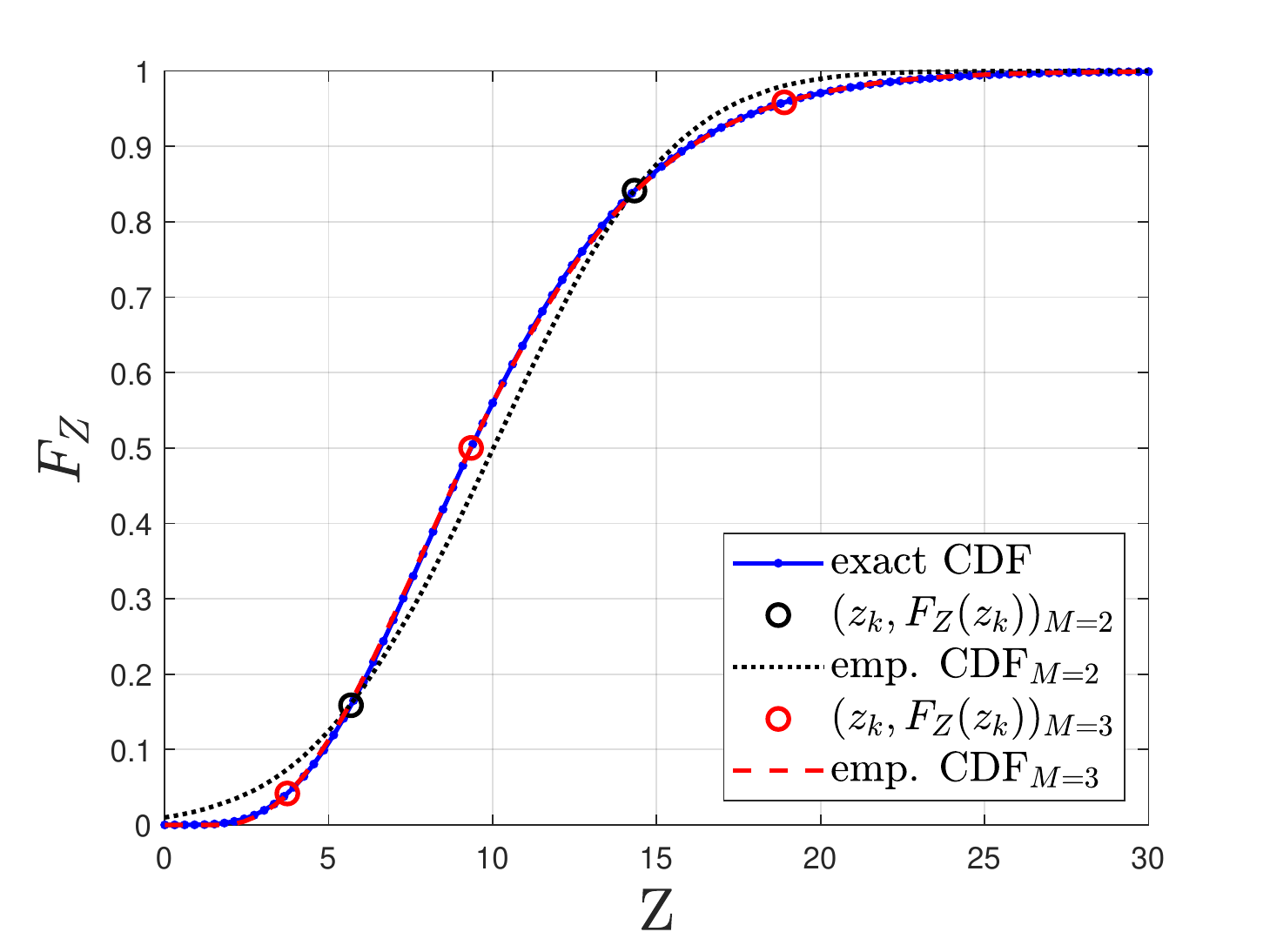} }}%
    \caption{\footnotesize Application of SCMC to $Z\sim \chi^2(\mathrm{df})$, $\mathrm{df}=10$. ``Cheap'' random variable $\xi\sim \N(0,1)$. Lagrange interpolation with $M=2,3$ collocation points (represented with black and red respectively, in blue, on the other hand, are given the exact curves).}
    \label{fig: Lag int M3}%
\end{figure}

At this stage, the interpolation rule $g_M$ is defined (remember that here we are assuming to use Lagrange interpolation), employing as reference values the $M$ pairs of collocation points $(\xi_k, z_k)$, $k=1,\dots,M$. We end up with an approximation of Equation (\ref{eq: identity sampling}), namely
\begin{equation}
    \label{eq: identity approx sampling}
    {Z}\overset{d}{\approx} g_M({\xi})=\sum_{i=0}^{M-1} \alpha_i \xi^i,
\end{equation}
where the coefficients of the Lagrange polynomial $\alpha_i=\alpha_i\big((\xi_k,z_k)_{k=1}^M\big)$, $i=1,\dots,M$, depend on the pairs of collocation points. For details about the error induced by the different interpolation rules and for the full computation in case of Lagrange interpolation, see \cite{GrWiSuOo}. In Figure \ref{fig: Lag int M3}a we can see the comparison between the map $g$ and its approximation $g_M$ (with $M=2,3$), while the effect of the approximation on the CDF $F_Z$ is illustrated in Figure \ref{fig: Lag int M3}b.

Eventually, a bullet-point algorithm to guide the implementation of the SCMC sampler is provided here.

\zbox{
{\bf Algorithm: SCMC sampler} \\
\label{alg: SCMC}
We consider the target (``expensive'') distribution $Z$, with CDF $F_Z(x)=\P[Z\leq x]$, and the ``cheap'' distribution $\xi$, with CDF $F_{\xi}(x)=\P[\xi\leq x]$.
\begin{enumerate}
\itemsep0.5em
\item Select $M\geq 2$ original collocation points $\xi_k$, $k=1,\dots,M$, for the ``cheap'' random variable $\xi$;
\item Evaluate the ``cheap'' CDF $F_\xi$ at the original collocation points, i.e. $F_{\xi}(\xi_k)$ for each $k=1,\dots,M$;
\item Invert the ``expensive'' CDF $F_Z$ to compute the pairs $(\xi_k, z_k)$, $k=1,\dots,M$, where $z_k:=F_Z^{-1}\big(F_\xi(\xi_k)\big)$;
\item Starting from the pairs of collocation points $(\xi_k, z_k)$, $k=1,\dots,M$, build the interpolating function $g_M$ (for instance, using Lagrange interpolation, $g_M(x):=\sum_{i=0}^{M-1}\alpha_i x^i$, with $\alpha_i=\alpha_i\big((\xi_k, z_k)_{k=1}^M\big)$, $i=0,\dots,M-1$);
\item Set $N$ as the desired number of samples to draw. For each $n=1,\dots,N$, sample $\Bar Z_n$ from $Z$ evaluating $g_M$ in the sample $\Bar\xi_n$ from $\xi$, i.e. $\Bar Z_n = g_M(\Bar\xi_n)$.
\end{enumerate}
}

\subsubsection{Stochastic Collocation Monte Carlo for time-integrated bridges}
\label{sssec: SCMC for Z}

This section is dedicated to summarize how this technique is employed in our framework, dealing with conditional time-integrated processes.

For a given $\tttheta$, namely for a given choice of model parameters ($\theta_0$) and time window ($t_1$ and $t_2$), let us consider as target distribution the conditional time-integrated process $Z(\theta|\cdot, \cdot)$. Moreover, let the two boundary conditions $(a,b)\in\R\times\R$ be given.
Once the set of original collocation points $\xi_k$, $k=1,\dots,M$, is specified, $M$ inversion of the CDF
\begin{equation}
    F(x|a,b):=\P\big[Z(\theta|a,b)\leq x\big],
\end{equation}
are performed to compute as many collocation points $z_k(a,b)$ through the relationship
\begin{equation}
\label{eq: identity explicit Z(t_1,t_2)}
    z_k(a,b)=F^{-1}(F_\xi(\xi_k)|a,b).
\end{equation}
The $M$ original collocation points $\xi_k$ together with the $M$ collocation points $z_k(a,b)$ are used to build the interpolating function $g_M(x|a,b)$, which means
\begin{equation}
    \label{eq: identity approx sampling Z(t1,t2)}
    {Z}(\theta|a,b)\overset{d}{\approx} g_M({\xi}|a,b)=\sum_{i=0}^{M-1} \alpha_i \xi^i,
\end{equation}
with $\alpha_i:=\alpha_i\big((\xi_k,z_k(a,b))_{k=1}^M\big)$, $i=1,\dots,M$.

Eventually, the sampling from $Z(\theta|a,b)$ can be performed evaluating $g_M(x|a,b)$ in the samples $\Bar{\xi}$ drawn from the ``cheap'' random variable $\xi$. We remark that a small number $M$ of collocation points\footnote{In principle, to build a $M-1$ degree Lagrange polynomial, we need $2M$ values (we have to store both the $\xi_k$ and the $z_k(a,b)$). In practice, the $\xi_k$ are chosen a priori, assuming we always know them.} contains the complete information needed to draw any number of samples from $Z(\theta|a,b)$, provided a suitable interpolation rule.

\section{Methodology description}
\label{sec: methodology}
In this section, we provide a detailed description of the methodology. We can recognize two separate phases: a one-time off-line stage, possibly computationally expensive, and an on-line stage, in which the actual sampling from the time-integrated bridge $Z(\theta|\cdot,\cdot)$ is performed. In Table \ref{tab: method scheme}, a scheme of the two phases is given.

\begin{table}[H]
\small
\begin{center}
\caption{Methodology scheme.}
\label{tab: method scheme}
\vspace{-0.05cm}
\resizebox{0.87\textwidth}{!}{\begin{tabular}{ll}
\hline\hline
\multicolumn{2}{c}{\multirow{2}{*}{OFF-LINE STAGE\qquad\qquad}}\\
\multicolumn{1}{c}{\multirow{2}{*}{Compression}}   &\multicolumn{1}{c}{\multirow{2}{*}{ANN training}} \\\\\hline
\multicolumn{1}{l|}{\begin{tabular}[c]{@{}l@{}}Given the tuple $\theta$, run MC\\ simulations; from MC paths\\ compute the grid of CPs.\end{tabular}}  & \multicolumn{1}{l}{\begin{tabular}[c]{@{}l@{}}For many randomly chosen $\theta$, compute the\\ CPs via compression to generate the training\\ set; train the Artificial Neural Network $H$.\end{tabular}}\\\\
\hline\hline
\multicolumn{2}{c}{\multirow{2}{*}{ON-LINE STAGE\qquad\qquad}}\\
\multicolumn{1}{c}{\multirow{2}{*}{ANN evaluation}}   &\multicolumn{1}{c}{\multirow{2}{*}{Decompression}} \\\\\hline
\multicolumn{1}{l|}{\begin{tabular}[c]{@{}l@{}}For a specific tuple $\theta$,\\ compute the corresponding\\ grid of CPs evaluating $H(\theta)$.\end{tabular}}  & \multicolumn{1}{l}{\begin{tabular}[c]{@{}l@{}}For specific boundary conditions $(a, b)$, ex-\\tract the collocation points by interpolation;\\ perform the sampling via SCMC sampler.\end{tabular}}
\end{tabular}}
\end{center}
\end{table}

\subsection{Off-line stage}
\label{ssec: off-line}
The goal of the off-line phase is to build a function $H$ which allows mapping any model parameters $\theta$ into a set of collocation points that can be employed for the sampling. As we saw in Section \ref{ssec: SCMC}, it is possible to store the complete information needed to sample from one single distribution in a small number of collocation points, that is $M$. Such a process that reduces the amount of information necessary during the MC simulation without a significant loss in accuracy is called ``compression'' and is fundamental for implementing the deep learning part. In the more general case, the information is stored in grids (see Section \ref{sssec: compr}), and the function $H$ ``learns'' how to map the parameters $\theta$ into the corresponding grids, then used for the sampling. Let us present now the details about the compression technique.

\subsubsection{Compression based on Monte Carlo samples}
\label{sssec: compr}
Let us recall that the target distribution is $Z(\theta|a,b)$ as defined in Equation (\ref{eq: integrated bridge definition}), whereas the cheap random variable $\xi$ is standard normal $\N(0,1)$. The goal is to sample from $Z(\theta|\cdot, \cdot)$ for any given pair of initial and final values $(a,b)\in \R\times \R$. This operation cannot be performed analytically. On the other hand, in principle, we can run an extremely precise classical Monte Carlo simulation to obtain an accurate empirical approximation of the target distribution. Nonetheless, the classical MC scheme has a low convergence speed. Consequently, to get reliable samples, a huge computational time is necessary, which is undesirable in applications. 

The CDF of $Z(\theta|a,b)$ varies smoothly with respect to the pair of boundary conditions $(a,b)$, and so does the map $g(\cdot|a,b)$ with the corresponding collocation points $z_k(a,b)$ (as defined in Equation (\ref{eq: identity explicit Z(t_1,t_2)})).
Therefore, to manage any boundary conditions $(a,b)$ we can compute the collocation points only for a discrete grid of boundary conditions and then recover the collocation points for a specific pair of boundary conditions by means of a suitable interpolation (see Section \ref{sssec: decompr}). In particular, the range of the initial realization and the range of the final realization is discretized in a few reference values, and for each pair of reference boundary conditions, a plain MC simulation\footnote{We recall that in general there is no analytic expression; hence we have to rely on MC simulation.} is performed off-line. The collocation points for the conditional time-integrated process are calculated for each pair of reference boundary conditions. At this point, for any choice of $(a,b)$ it is possible to recover by interpolation the corresponding collocation points for the conditional time-integrated process, and so the sampling from $Z(\theta|a,b)$ can be run employing SCMC sampler (Section \ref{sssec: SCMC for Z} and Algorithm \ref{alg: SCMC}).

\begin{figure}[t!]%
    \centering
    \subfloat[\centering]{{\includegraphics[width=7.cm]{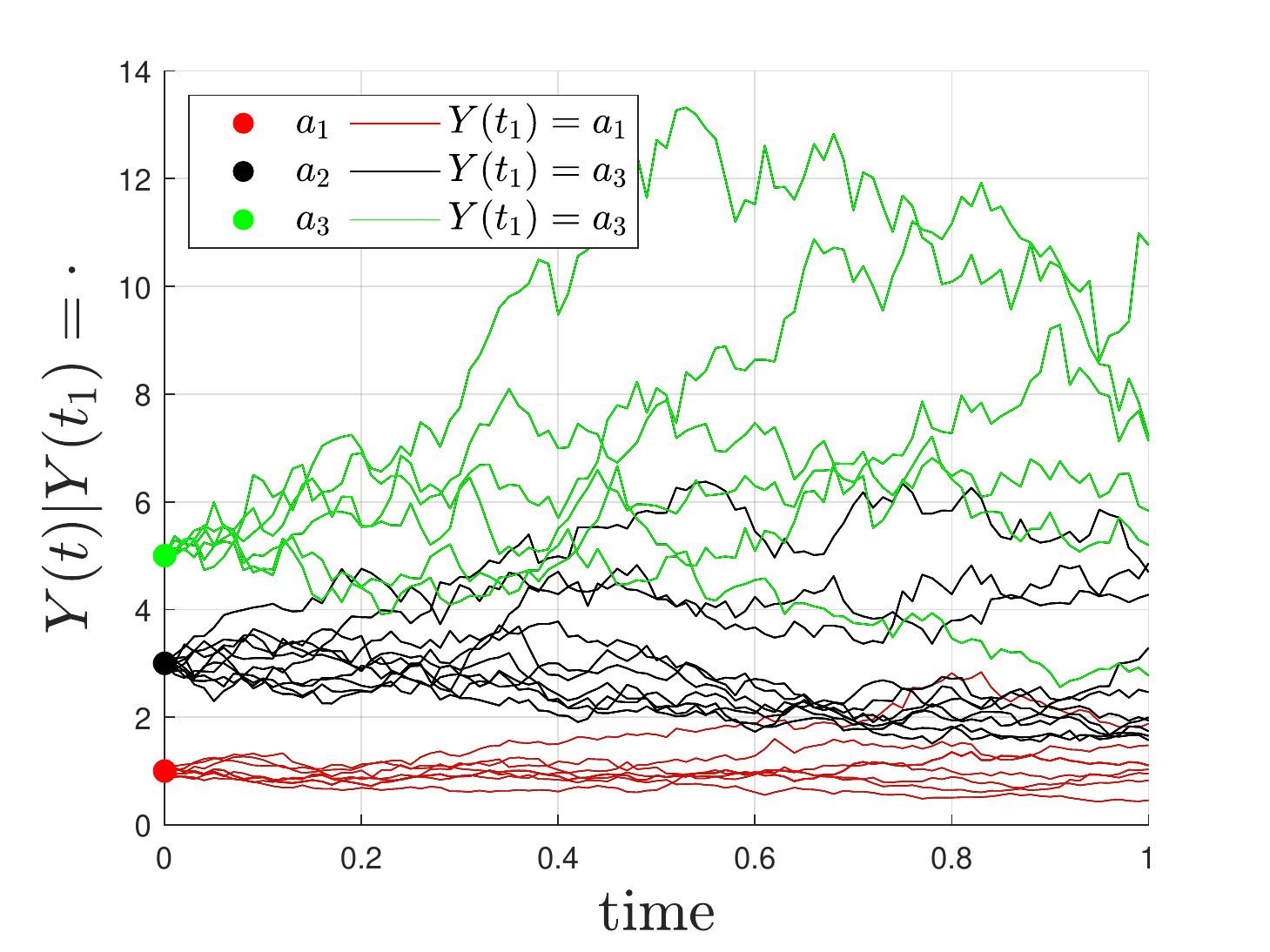} }}%
    ~
    \subfloat[\centering]{{\includegraphics[width=7.cm]{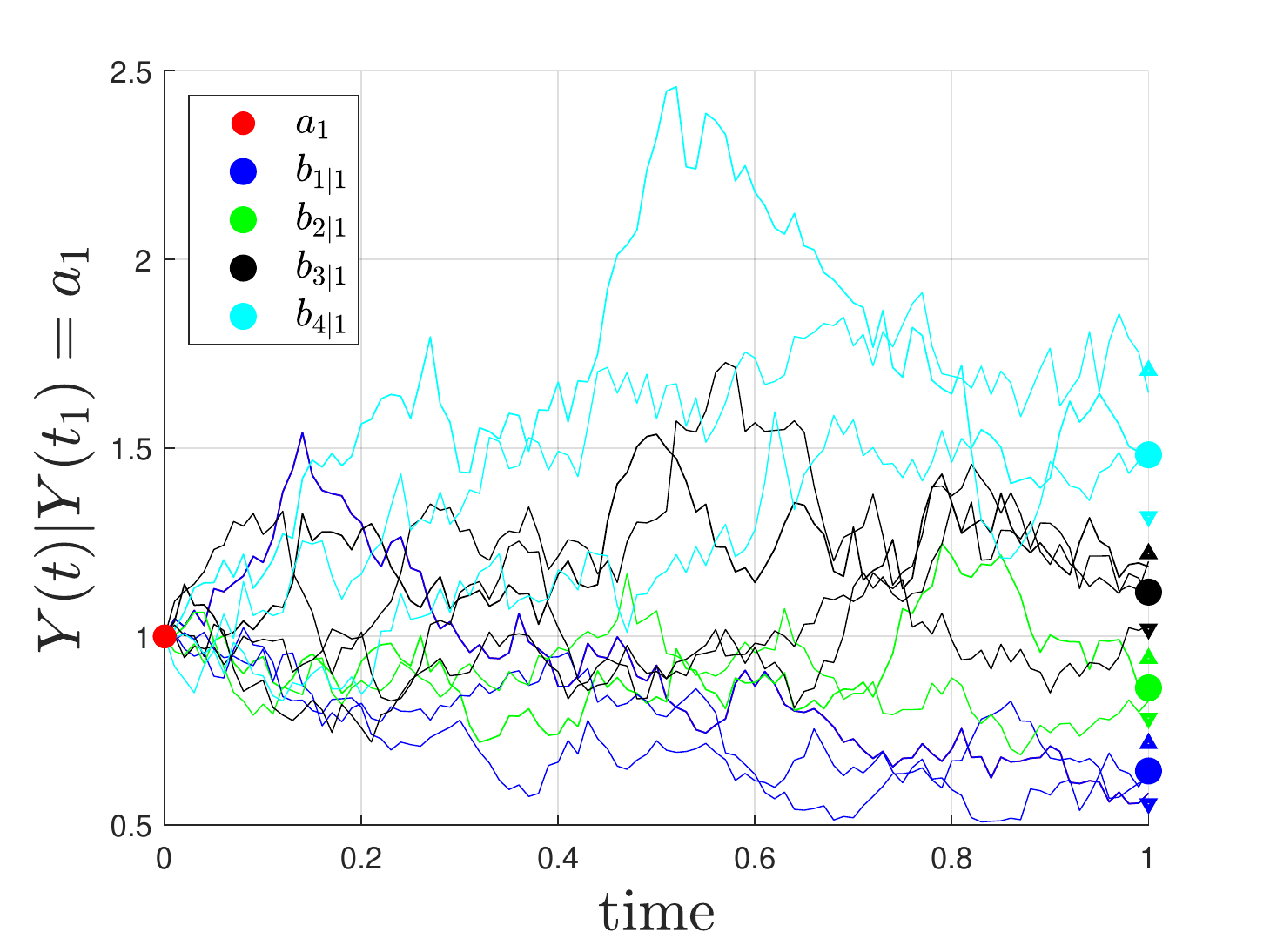} }}%
    \caption{\footnotesize Left: some Monte Carlo paths for a GBM given different initial values $a_i$, $i=1,2,3$ (respectively in red, black and green). Right: some Monte Carlo paths starting at $a_1$, divided according to the final value $b_{h|1}$, $h=1,2,3,4$ (respectively in blue, green, black and cyan).}%
    \label{fig: illustration discretizations}%
\end{figure}

The compression procedure is implemented as follows.
Firstly, the range of the random variable $Y(t_1)$ is discretized into the grid $A$ of length $M_a\geq 2$ (see Figure \ref{fig: illustration discretizations}a), that is
\begin{equation}
    A:=\begin{bmatrix}
    a_1&\dots& a_i&\dots& a_{M_a}
    \end{bmatrix}.
\end{equation}
A simple yet effective criterion to build $A$ is with equally spaced values, namely
\begin{equation}
    a_{i+1}-a_{i}=a_{i}-a_{i-1}, \qquad \forall i=2,\dots,M_a-1,
\end{equation}
nonetheless other more refined criteria exist (see Section \ref{sssec: CIR}).

Then, for each reference initial value $a_i$, $i=1,\dots,M_a$, we consider the conditional random variable at final time $Y(t_2)|Y(t_1)=a_i$ (see Figure \ref{fig: illustration discretizations}b). The range of each (conditional) random variable is discretized into $M_b\geq 2$ points. The criterion to perform this discretization is different from the one employed for $A$. Indeed, it is convenient to divide the range in bins in such a way the probability of one path ending in each bin is the same, namely for each $a_i$, $i=1,\dots,M_a$, the range of $Y(t_2)|Y(t_1)=a_i$ is decomposed into bundles
\begin{equation}
    \big(b_-, b_{1|i}\big],\qquad
    \big(b_{h|i}, b_{h+1|i}\big],\:\: h=1,\dots,M_{b}-1, \qquad
    \big(b_{M_b|i}, b_+\big),
\end{equation}
where for each $h=0,\dots,M_b$ (and $i=1,\dots,M_a$) it holds
\begin{equation}
    \mathbb{P}\Big[Y(t_2)\in\big(b_{h|i}, b_{h+1|i}\big]\Big| Y(t_1)=a_i\Big]=\frac{1}{M_b+1},
    \label{eq: bundle criterium}
\end{equation}
with the end points $b_{0|i}=b_-$, $b_{M_b+1|i}=b_+$ and the open interval if $h=M_b$. In other words, the (conditional) discretization for the final value is taken as an equally spaced quantile grid of the distribution of $Y(t_2)|Y(t_1)=a_i$ for each $i=1,\dots,M_a$.
Notice that in general the left-end point and the right-end point, $b_{-}$ and $b_+$, depend on the structural parameters of the process $Y(t)$, but not on the initial value $a_i$. Moreover, they are not needed in the application of the technique, hence they are not stored. 

\begin{figure}[t]%
    \centering
    \subfloat[\centering]{{\includegraphics[width=7.cm]{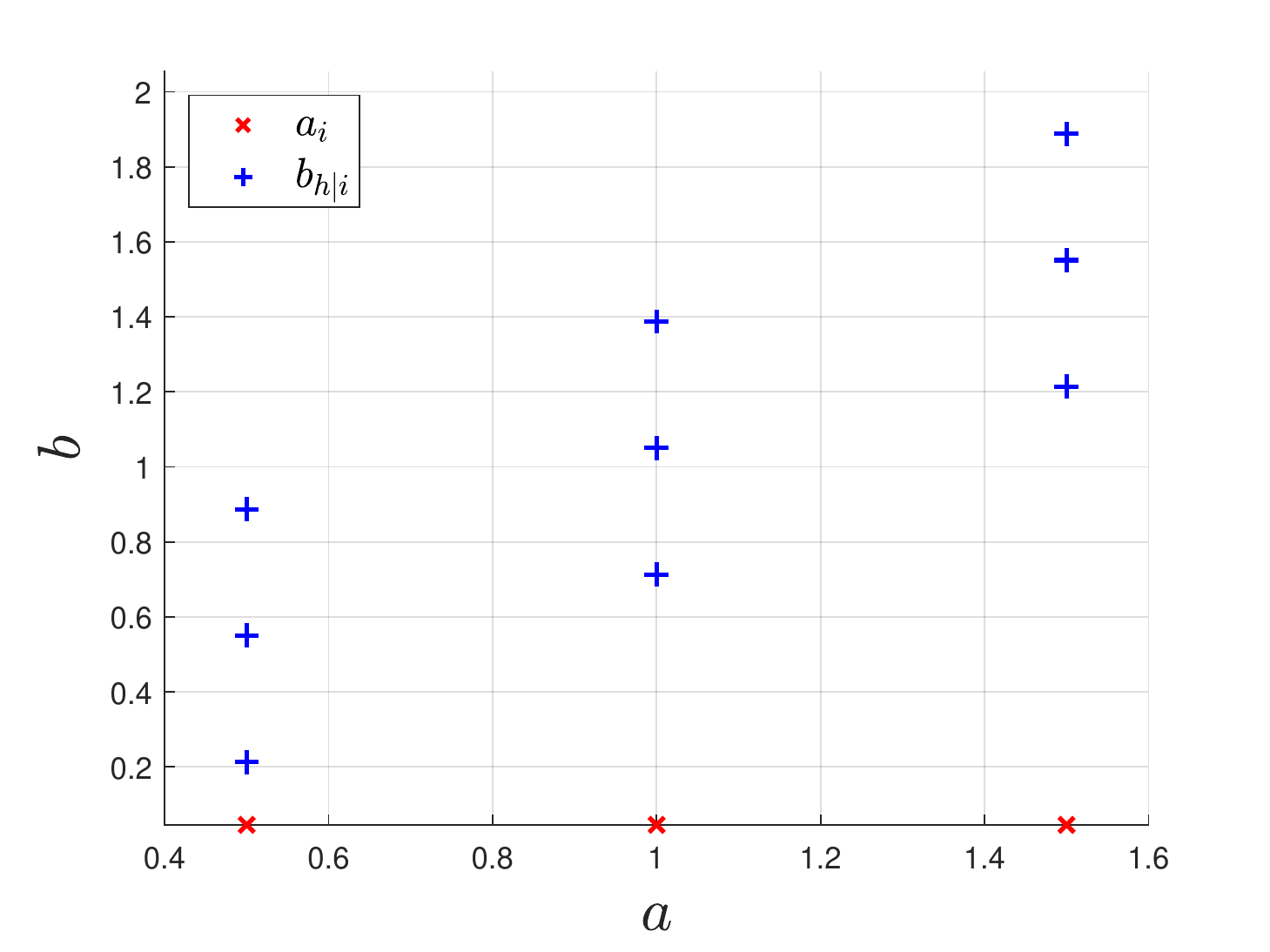} }}%
    ~
    \subfloat[\centering]{{\includegraphics[width=7.cm]{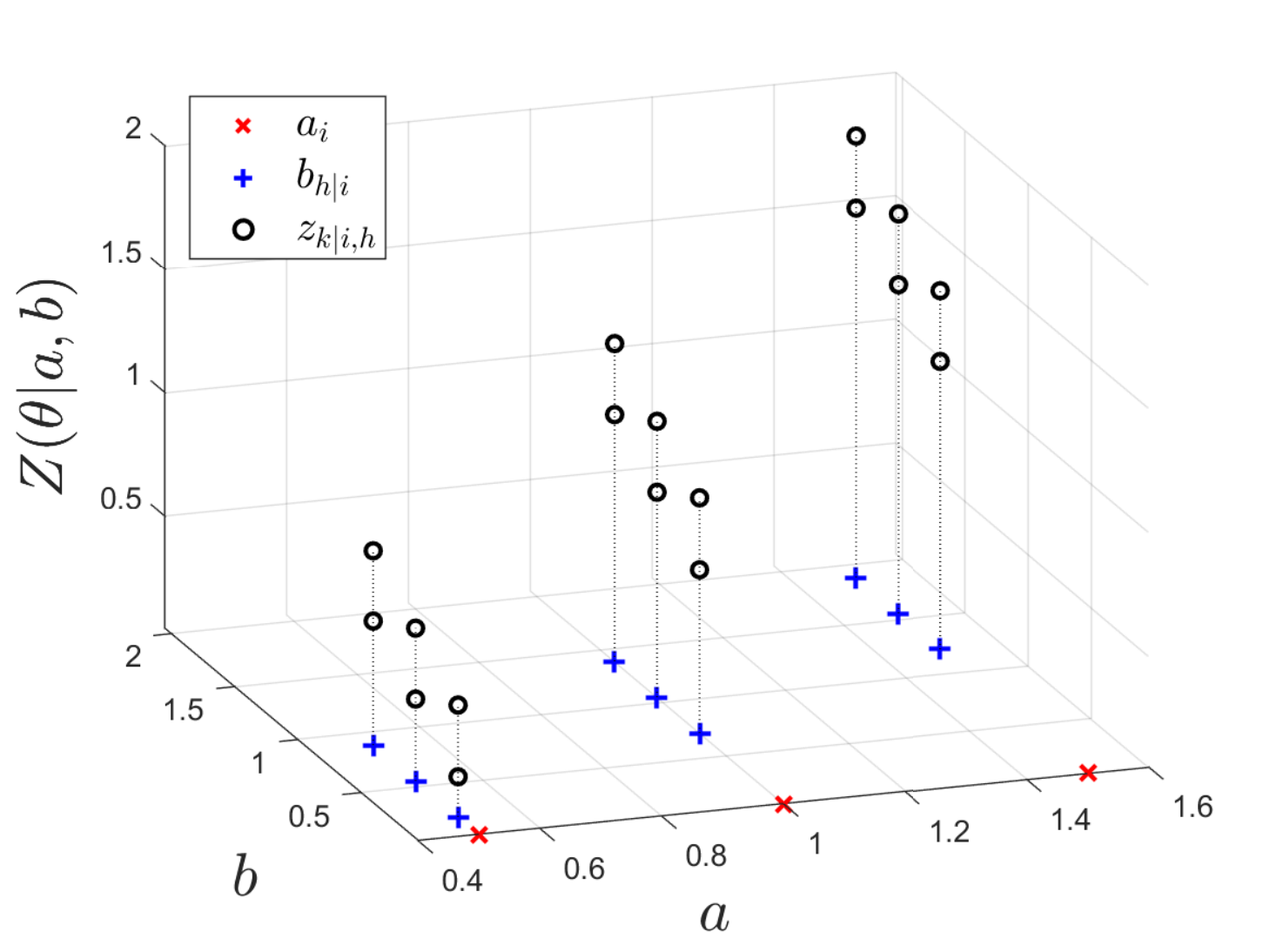} }}%
    \caption{\footnotesize Example of grids $B$ (left) and $C$ (right) for conditional time-integrated ABM: $M_a=3$, $M_b=3$, $M=2$.}%
    \label{fig: grids B and C}%
\end{figure}

The result of this operation is the 2D grid of the reference final values conditioned on the reference initial values $a_i$, $i=1,\dots,M_a$, which can be represented through the $M_a\times M_b$ matrix (see Figure \ref{fig: grids B and C}a)
\begin{equation}
    B:=\begin{bmatrix}
    b_{1|1}&\dots& b_{h|1}&\dots & b_{M_b|1}\\
    \vdots & & \vdots && \vdots\\
    b_{1|i}&\dots& b_{h|i}&\dots & b_{M_b|i}\\
    \vdots & & \vdots && \vdots\\
    b_{1|M_a}&\dots& b_{h|M_a}&\dots & b_{M_b|M_a}
    \end{bmatrix}_{M_a\times M_b},
\end{equation}
where the $i^{th}$ row corresponds to the conditional discretization for the final value of the process given as initial value $a_i$.

The two grids $A$ and $B$ are employed as reference values for the initial value $a$ and the final value $b$. Particularly, at each pair $(a_i, b_{h|i})$, $i=1,\dots,M_a$, $h=1,\dots,M_b$, corresponds the conditional time-integrated process $Z(\theta|a_i, b_{h|i})$. The distribution of $Z(\theta|a_i, b_{h|i})$ is compressed into the $M\geq 2$ collocation points as shown in Section \ref{sssec: SCMC for Z}. In practice, the set of original collocation points $\xi_k$, $k=1,\dots,M$,
\begin{equation}
    \begin{bmatrix}
        \xi_{1}&\dots & \xi_{k} &
        \dots &
        \xi_{M}
    \end{bmatrix},
    \label{eq: original CPs}
\end{equation}
is specified a priori (we use the same \emph{optimal}\footnote{Using the same terminology as in \cite{GrWiSuOo}, the $M$ \emph{optimal} original collocation points are defined as the roots of the $M-1$-degree orthogonal polynomial (for details see \cite{GoWe, GrWiSuOo}).} one for each pair of reference boundary conditions ($a_i, b_{h|i}$)). Then, for each $Z(\theta|a_i, b_{h|i})$, the $M$ collocation points $z_{k|i,h}:=z_k(a_i,b_{h|i})$, $k=1,\dots,M$, are computed as numerical quantiles resulting from the expensive plain Monte Carlo simulation employed to reproduce the distribution of the reference time integrated bridge $Z(\theta|a_i, b_{h|i})$. The whole set of $M_a\times M_b \times M$ collocation points can be stored into the 3D matrix (see Figure \ref{fig: grids B and C}b)
\begin{equation}
    C=\big[z_{k|i,h}\big]_{i,h,k},\qquad
    i=1,\dots,M_a,\quad
    h=1,\dots,M_b,\quad
    k=1,\dots,M.
\end{equation}

\begin{rem}[Numerical computation of matrix $C$]
\label{rem: numerical computation C}
Notice that in general, given $a_i$ and $b_{h|i}$, the computation of the corresponding collocation points $z_{k|i,h}$ cannot be performed analytically. This means that we have to relay on numerical approximation via MC simulation. The idea is to exploit the empirical distribution of $Z(\theta|a_i,N_{h|i})$, with $N_{h|i}=(b_{h|i}-\epsilon_d, b_{h|i}+\epsilon_u)$, as an approximation of the distribution of $Z(\theta|a_i, b_{h|i})$. On one hand the approximation is more precise the smaller are $\epsilon_d$ and $\epsilon_u$; on the other hand, the smaller is the neighborhood $N_{h|i}$ the less stable is the empirical distribution of the (approximated) conditional integral obtained from the MC simulation (assuming the number of overall paths remains unchanged).
\end{rem}

The matrices $A$, $B$ and $C$ are the result of the compression part. Indeed, the reference values in them will allow to recover the MC samples from the distribution $Z(\theta|a,b)$ for any choice of initial and final values, namely for any choice of boundary conditions $(a,b)\in \R\times \R$. 
Nonetheless, given a set of parameters $\theta$, building the corresponding reference grids -- $A$, $B$, $C$ -- may require some time. To speed-up this computation, we reformulate the problem as a regression one and we tackle it employing deep learning techniques presented in the next section. We conclude this part with the algorithm of the compression phase.

\zbox{
{\bf Algorithm: Compression}
\label{alg: compression}\\
Let the set of parameters $\theta$ and the range $[a_{min}, a_{max}]$ for the initial value be given. Let the dimensions of the grids be specified by $M_a\geq 2$, $M_b\geq 2$ and $M\geq 2$.

\begin{enumerate}
\itemsep0.5em
\item Generate the equally-spaced grid $A$ as 
\begin{equation*}
    a_i = a_{min} + (i - 1)\cdot\delta_a, \qquad \delta_a:=\frac{a_{max} - a_{min}}{M_a-1}, \quad i = 1, \dots, M_a;
\end{equation*}
\item For each $i=1,\dots,M_a$, analytically generate the $i^{th}$ row of the grid $B$ as the $M_b$ quantiles $b_{h|i}$ ($h=1,\dots,M_b$) of the distribution of $Y(t_1+t_2)|Y(t_1)=a_i$. The quantiles are taken in such a way that they span equiprobable bundles on the range of the variable $Y(t_1+t_2)|Y(t_1)=a_i$;
\item For each $i=1,\dots,M_a$, run a MC simulation starting at $a_i$;
\item Given the $M$ original CPs $\xi_k$ ($k=1,\dots,M$), for each $i=1,\dots,M_a$ and $h=1,\dots,M_b$, compute the $M$ CPs $z_{k|i,h}$ as quantiles of the empirical distribution of $Z(\theta|a_i,N_{h|i})$ (Remark \ref{rem: numerical computation C}). Collect them in the 3D matrix $C$.
\end{enumerate}
}
\subsubsection{Artificial Neural Network training}
\label{sssec: ANN}
This part aims to speed up the computation of the grids $A$, $B$ and $C$ during the on-line phase. Of course, this has a cost, but it is paid during the off-line phase and does not affect the efficiency of the method during applications. 

In the same fashion as in \cite{LiGrOo}, for any choice of model parameters $\theta$, the computation of the grid $C$ is given in terms of a regression problem. An artificial neural network is trained to map $\tttheta$ into the collocation points stored in $C$. The net's training is performed using an artificial training set obtained via accurate MC simulation. Observe that the grid $A$ is specified a priori and depends on the particular application, whereas the grid $B$ is usually attainable analytically. This is because the reference (conditional) final values in $B$ are nothing but quantiles of $Y(t_2)|Y(t_1)$, which often can be described in terms of well-known distributions (for instance in the applications in Section \ref{sec: appl}). Hence, there is no need to train the ANN to learn $A$ and $B$, but it is enough to train it to map $\theta$ into $C$.

The learning task, here, is tackled employing a fully-connected neural network. Such a map can be represented as the composition function
\begin{equation}
            H(\tttheta|\W)=h^{(L)}(\dots h^{(2)}(h^{(1)}(\tttheta);\w^{(1)},\b^{(1)});\w^{(2)},\b^{(2)})\dots;\w^{(L)},\b^{(L)}),
        \end{equation}
where $L$ is the numbers of layers and $\W=\{\w^{(1)},\b^{(1)},\w^{(2)},\b^{(2)},\dots,\w^{(L)},\b^{(L)}\}$ with $\w^{(\ell)}$ the matrix of weights and $\b^{(\ell)}$ the vector of biases of the $\ell^{th}$ layer. Each map $h^{(\ell)}$ is an inner product between the output $\u^{(\ell-1)}$ from the previous layer and the weights matrix (plus a bias term) composed with a nonlinear activation function $\phi^{(\ell)}$, namely the $j^{th}$ component of the $\ell^{th}$ layer's output vector is given as
\begin{equation}
    \u^{(\ell)}_j = \phi^{(\ell)}\bigg(\sum_{i=1}^{N_{\ell-1}} w_{i,j}^{(\ell)}\u_i^{(\ell-1)} + \b_j^{(\ell)}\bigg),\qquad j=1,\dots,N_{\ell},
\end{equation}
where $N_{\ell}$ is the number of neurons in the $\ell^{th}$ layer.

Once the hyper-parameters of the ANN are specified, the net is trained on the so-called \emph{training set}, indicated with $\boldsymbol{T}$. Each element in $\boldsymbol{T}$ is an (input-output) pair made of the model parameters $\theta$ and the corresponding collocation points $C$. A user-defined loss function $J(\W)$ is selected to measure the ``distance'' between the real output -- the second entry of each element in $\boldsymbol{T}$ - and the predicted one -- the one obtained evaluating the net at the first entry. The training of the ANN consists in minimizing the function $J(W|\boldsymbol{T})$ with respect to the weights and biases stored in $\W$. 
Hence, the optimal weights and biases are obtained as
\begin{equation}
    W_{op}=
    \underset{\W}{\text{argmin }} J(\W|\boldsymbol T).
\end{equation}
A minor part of the training set $\boldsymbol{T}$ goes under the name of \emph{test set} and is not used for the actual training, but only to measure the accuracy of the calibrated ANN on unseen data. Finally, for what concerns the remaining part of $\boldsymbol{T}$, again a minor part is selected and used as \emph{validation set}, namely a set of data employed to avoid over-fitting of the ANN to the training set itself (we want the network to work for any choice of model parameters $\theta$ and not only for the ones present in the training set).
        
The training phase may require a considerable amount of time. The reasons are mainly two. First, the artificial training set $\boldsymbol{T}$ is built through plain MC simulations. Hence, to get a satisfactory training set, we need to repeat a really accurate MC simulation a huge number of times. Secondly, in principle, the actual optimization process is a problem that involves a huge number of dimensions, so it can be a challenging operation to perform. A usual approach is to optimize the weights (and biases) through \emph{back-propagation} using stochastic gradient descend.
On the other hand, all the training phase is performed off-line, and it does not affect the on-line part of the method (described in the next section), which remains extremely fast.

\subsection{On-line stage}
\label{ssec: online}
Here, we inspect how the actual simulation is performed according to the proposed method. We can identify two steps: given the set of model parameters $\theta$, the reference collocation points -- stored in $C$ -- are computed employing the ANN; then, the boundary conditions $(a,b)$ are used to obtain the collocation points for $Z(\theta|a,b)$ and SCMC is applied for the sampling.

\subsubsection{Fast collocation points computation}
\label{sssec: CP computation}
The reason why we accept a huge computational time during the off-line stage is stated here. Indeed, for what we have seen so far, given a set of model parameters $\theta$, the construction of the grid of collocation points $C$ is based on MC simulation, and so it takes a considerable amount of time. On the other hand, the evaluation of an ANN is extremely fast. A properly trained neural network is the tool we can exploit to speed up the computation of the collocation points for \emph{any} choice of model parameters $\theta$. Hence, as a first step of the on-line phase, we evaluate the ANN $H$ -- resulting from the off-line stage -- at the set of model parameters $\theta$. The result is the collocation points stored in the grid $C$, namely
\begin{equation}
    \tttheta\longmapsto H(\tttheta)=C,\quad \tttheta\in \Theta_{H},
\end{equation}
where $\Theta_{H}\subset \Theta$ is the subset of the model parameters domain in which the ANN $H$ performs accurately. In general, $\Theta_H\subset\Theta_T$, with $\Theta_T$ the subset of the model parameters domain on which the training set $\boldsymbol{T}$ was built. Empirical evidence shows that approaching the boundaries of $\Theta_T$, the accuracy of the network deteriorates, whereas the ANN is reliable ``far'' from the boundaries (but inside the training domain). 

\subsubsection{Decompression and sampling}
\label{sssec: decompr}
At this point, we can assume the three grids of collocation points $A$, $B$, and $C$ to be known. Indeed, $A$ is specified a priori, $B$ is usually attainable analytically, and $C$ can be computed employing the artificial neural network $H$. The purpose of this section is to show how to employ this compact amount of information (in terms of memory storage) to produce fast and accurate MC simulations, no matter what the boundary conditions $a\in \R$ and $b\in \R$ are. Therefore, we call this stage ``decompression''.

Let $(a,b)$ be any pair of feasible real values for the process $Y(t)$ at initial time $t_1$ and at final time $t_2$. As shown in Section \ref{sssec: SCMC for Z}, the $M$ original collocation points $\{\xi_k\}_{k=1}^M$ and the $M$ collocation points $\{z_k(a,b)\}_{k=1}^M$ can be employed to build the interpolating function $g_M(x|a,b)$ and then the map $g_M$ is used for the sampling.
Nonetheless, in general the elements of the pair $(a,b)$ are not available in the grids $A$ and $B$. Hence, also the corresponding collocation points $z_k(a,b)$, $k=1,\dots,M$, are not directly available from $C$. The most natural solution is to employ the pair $(a,b)$ to enter into the 3D matrix $C$ and perform a suitable interpolation on the ``neighbour'' collocation points available in $C$. This allows to recover (approximated) collocation points $\Tilde{z}_k(a,b)$, $k=1,\dots,M$.

Therefore, we can recognize two stages in the decompression phase. Firstly, it is necessary to recover the approximated collocation points $\Tilde{z}_k(a,b)$ corresponding to the pair of boundary conditions $(a,b)$, namely
\begin{equation}
    \Tilde{z}_k(a,b)=\psi_k(a, b; A, B, C), \quad k=1,\dots,M,
\end{equation}
where $\psi_k$, $k=1,\dots,M$, are the approximating maps which performs the interpolation, taking as input the values $a$ and $b$ and using them to enter in the grid $C$.
Then, the collocation points $\Tilde{z}_k(a,b)$ are employed to build the interpolating function $g_M(x|a,b)$, which is eventually used into the SCMC sampler, as described in Section \ref{sssec: SCMC for Z}.

The specification of the maps $\psi_k$, $k=1,\dots,M$, is strongly dependent on the underlying process (see Section \ref{ssec: models}). For what concerns our applications, inspecting the structure of the matrix $C$, good results are achieved with linear interpolation within the ranges covered by $A$ and $B$. On the other hand, most of the times when extrapolation is needed, namely when the values of interest $(a,b)$ are outside the ranges covered by $A$ and $B$, better results are obtained employing quadratic interpolation (which becomes extrapolation outside the ranges of $A$ and $B$).

Since the grid $B$ is filled with the \emph{conditional} final values $b_{h|i}$'s (conditional with respect to the $a_i$'s stored in $A$), the maps $\psi_k$'s have to perform two consequent 1D interpolations (or extrapolations). The first one along the direction of the initial value $a$  -- say ``$a$-direction'' -- and the second one, performed on the resulting values from the first one, is done along the direction of the final value $b$ -- the ``$b$-direction''. The maps $\psi_k$'s are nothing but the composition of the two interpolations, first along the ``$a$-direction'' and later along the ``$b$-direction''.

Once the approximated collocation points are available, the sampling is a straightforward application of the SCMC technique. Indeed, it is enough to employ for the sampling Equation (\ref{eq: identity approx sampling Z(t1,t2)}) using as collocation points the $\Tilde{z}_k(a,b)$, namely
\begin{equation}
    Z(\theta|a,b)\overset{d}{\approx}{g}_{M}(\xi|a,b),
\end{equation}
where ${g}_M$ is the interpolation rule spanned by the approximated collocation points $\{\Tilde{z}_{k}(a,b)\}_{k=1}^{M}$ together with the original CPs $\{\xi_k\}_{k=1}^{M}$. The sampling now is a simple evaluation of the interpolation rule $g_M(x|a,b)$ at a set of values obtained as samples from the ``cheap'' random variable $\xi$ (in our case, standard normal).

In the case of Lagrange interpolation, we end up with
\begin{equation}
    {g}_{M}(x|a,b):=\sum_{i=0}^{M-1} \alpha_i x^i,
\end{equation}
where $\alpha_i=\alpha_i\big((\xi_k,\Tilde{z}_k(a,b))_{k=1}^{M}\big)$, $i=0,\dots,M-1$, are obtained as a linear transformation of the $\Tilde{z}_k(a,b)$'s. In particular the coefficients $\alpha_i$'s are the solution of the linear system
\begin{equation}
\label{eq: CP from coefficients}
    \begin{bmatrix}
        1 & \xi_1 & \xi_1^2 & \cdots & \xi_1^{M-1} \\
        1 & \xi_2 & \xi_2^2 & \cdots & \xi_2^{M-1} \\
        \vdots & \vdots & \vdots & \cdots & \vdots \\
        1 & \xi_M & \xi_M^2 & \cdots & \xi_M^{M-1} \\
    \end{bmatrix}
    \begin{bmatrix}
        \alpha_0 \\
        \alpha_1 \\
        \vdots \\
        \alpha_{M-1} \\
    \end{bmatrix}
    =
\begin{bmatrix}
    \tilde z_1(a,b)\\
    \tilde z_2(a,b)\\
    \vdots\\
    \tilde z_M(a,b)\\
\end{bmatrix}.
\end{equation}
Such a solution exists and it is unique, if the Vandermonde matrix in Equation (\ref{eq: CP from coefficients}) is built employing the \emph{optimal} original collocation points. The bijective linear relationship between the coefficients and the collocation points will be extremely useful during application, allowing fast sampling which otherwise could not be guaranteed.

The meaning of decompression is now clear. Indeed, starting from a relatively small number of values, stored into the grids $A$, $B$, $C$, we are able to decompress this information to sample from the conditional time-integrated process $Z(\theta|a,b)$ for any choice of initial and final realizations $(a,b)$, performing only few ``cheap'' operations from the computational perspective.

We conclude this part wrapping up the whole procedure described so far in a bullet-point algorithm.

\zbox{
{\bf Algorithm: On-line procedure}
\label{alg: procedure}

Let the set of model parameters $\tttheta$ and the ANN $H$ be given. Assume further that are given $N$ pairs of different boundary conditions $(a^{(n)},b^{(n)})$ corresponding to the $N$ desired samples, each one from the different distribution $Z(\theta|a^{(n)}, b^{(n)})$.

\begin{enumerate}
\itemsep0.5em
\item Evaluate $H$ in $\tttheta$ to get the grids $B$, $C$ ($A$ is already known);

\item (For each $n=1,\dots,N$) Compute the approximated collocation points 
\begin{equation*}
    \Tilde{z}_k(a^{(n)},b^{(n)})=\psi_k(a^{(n)}, b^{(n)}; A, B, C),\qquad k=1,\dots,M:
\end{equation*}
$\psi_k$'s perform 2D interpolation (first along the ``$a$-direction'', then along the ``$b$-direction'') on the grid $C$. The kind of interpolation may depend on the specific process;

\item (For each $n=1,\dots,N$) Define the SCMC interpolating map $g_M(x|a^{(n)},b^{(n)})$ on the pairs of collocation points $(\xi_k,\Tilde{z}_k(a^{(n)},b^{(n)}))$, $k=1,\dots,M$ (different choices available as Lagrange interpolation, Chebyshev interpolation, \dots);

\item (For each $n=1,\dots,N$) Draw $\Bar\xi_n$ from the ``cheap'' random variable $\xi$ (for instance, standard normal);

\item (For each $n=1,\dots,N$) Evaluate the interpolating map $g_M(x|a^{(n)},b^{(n)})$ in $\Bar{\xi}_n$ to get the desired sample from $Z(\theta|a^{(n)},b^{(n)})$, i.e.
\begin{equation*}
    \big(\Bar Z(\theta|a^{(n)},b^{(n)})\big)_n=g_M(\Bar\xi_n|a^{(n)},b^{(n)}).
\end{equation*}
\end{enumerate}

Observe that, since many computing tools (available for instance in \textsc{Matlab}, or in the \texttt{numpy} library of \textsc{Python}) allow parallelization, the $N$ operations (one for each pair $(a^{(n)}, b^{(n)})$) in each bullet-point 2-5 can be performed in parallel achieving a reduced computational time.
}

\subsection{Discussion on the errors}
\label{ssec: error}
In this section, we briefly discuss the errors introduced by the methodology proposed. There are four different stages in which we introduce some kind of error. First, the application of SCMC (see Section \ref{ssec: SCMC}) induces an error due to the substitution of the map $g$ in Equation (\ref{eq: identity explicit}) with the interpolating function $g_M$; second, in the ``compression'' phase (see Section \ref{sssec: compr}) the computation of the collocation points $z_{k|i,h}$ is performed numerically considering all the processes with final value into a (small) neighborhood of the actual final condition $b_{h|i}$ (see Remark \ref{rem: numerical computation C}); moreover, in the ``decompression'' phase (see Section \ref{sssec: decompr}) the collocation point $\Tilde{z}_{k}(a,b)$ are obtained through interpolation or extrapolation between the reference values stored in $C$; eventually, there is the error due to the usage of the ANN to map the parameters $\theta$ into the corresponding collocation points $C$ (see Section \ref{sssec: CP computation}).

\subsubsection{SCMC error}
The \emph{optimal} choice of the cheap CPs $\xi_k$ (combined with the fact that $\xi\sim\N(0,1)$) allows to connect the error associated to the SCMC technique with that due to computing an integral with the Gauss-Hermite quadrature rule. 

Using the simpler notation of Section \ref{ssec: SCMC}, if we assume the map $g$ derivable at least $M$ times in the interval $I_\xi:=(\xi_1,\xi_M)$, one can easily estimate the \emph{absolute distance} $\epsilon_M^{(g)}$ between $g$ and $g_M$. It is given by the classical error induced by Lagrange interpolation, i.e., 
\begin{equation}
\label{eq: error for g}
    \epsilon_M^{(g)}(x):=\big|g(x)-g_M(x)\big|=\bigg|\frac{1}{M!}\frac{\d^M g(x)}{\d x^M}\Big|_{x=\eta_1}\prod_{k=1}^{M}(x-\xi_k)\bigg|,
\end{equation}
with $\eta_1\in I_\xi$. Clearly $\epsilon_M^{(g)}$ can be controlled, taking as $\eta_1$ the value that maximizes the $M^{th}$ derivative of $g$ on $I_\xi$.
Observe that the more the map $g$ is \emph{similar} to a polynomial, of degree at most $M-1$, the smaller the error $\epsilon_M^{(g)}$ is. In particular, if $g$ is \emph{exactly} a polynomial (of degree at most $M-1$), the error is zero.

More interesting from a probabilistic point of view is the mean square error $ \epsilon_M^{(MSE)}$ between the exact distribution $g(\xi)$ and that obtained via SCMC $g_M(\xi)$. To deduce $\epsilon_M^{(MSE)}$ we heavily exploit the \emph{optimal} choice of the cheap CPs $\xi_k$ and the standard normal distribution of $\xi$. Indeed, by definition
\begin{equation}
\label{eq: GaussHermite quad}
    \epsilon_M^{(MSE)}:=\E[(g(\xi)-g_M(\xi))^2]=\int_\R (g(x)-g_M(x))^2 f_\xi(x)\d x,
\end{equation}
with $f_\xi$ probability density function (PDF) of $\xi\sim\N(0,1)$. The last term in the Equation (\ref{eq: GaussHermite quad}) can be connected -- after appropriate rescaling (see \cite{GrWiSuOo}) -- to the computation of an integral with the Gauss-Hermite quadrature rule (see \cite{GoWe}) built on the \emph{optimal} CPs $\xi_k$, i.e.,
\begin{equation}
    \int_\R (g(x)-g_M(x))^2 f_\xi(x)\d x=\sum_{k=1}^M (g(\xi_k)-g_M(\xi_k))^2 \omega_k + \epsilon_M^{(GH)} = \epsilon_M^{(GH)},
\end{equation}
where $\omega_k$ are appropriate \emph{weights} and $\epsilon_M^{(GH)}$ is the error due to the computation of the integral via the Gauss-Hermite quadrature rule. We observe that the last equality is a trivial consequence of the fact that the interpolating function $g_M$ is constructed from the pairs $(\xi_k, z_k)$, with $z_k:=g(\xi_k)$, therefore $g_M(\xi_k)=z_k=g(\xi_k)$ for every $k=1,\dots,M$. Finally, the error $\epsilon_M^{(GH)}$ is in general known in the literature (see \cite{AbSt}) and from it we deduce that the mean square error $\epsilon_M^{(MSE)}$ has the form
\begin{equation}
    \epsilon_M^{(MSE)} = \epsilon_M^{(GH)} = \frac{M!\sqrt{\pi}}{2^M}\frac{\Psi^{(2M)}(\eta_2)}{(2M)!},
\end{equation}
with $\Psi(x):=\big(g(x)-g_M(x)\big)^2$ and $-\infty<\eta_2<+\infty$.

\subsubsection{Compression error}
\begin{figure}[b!]%
    \centering
    \subfloat[\centering]{{\includegraphics[width=7.cm]{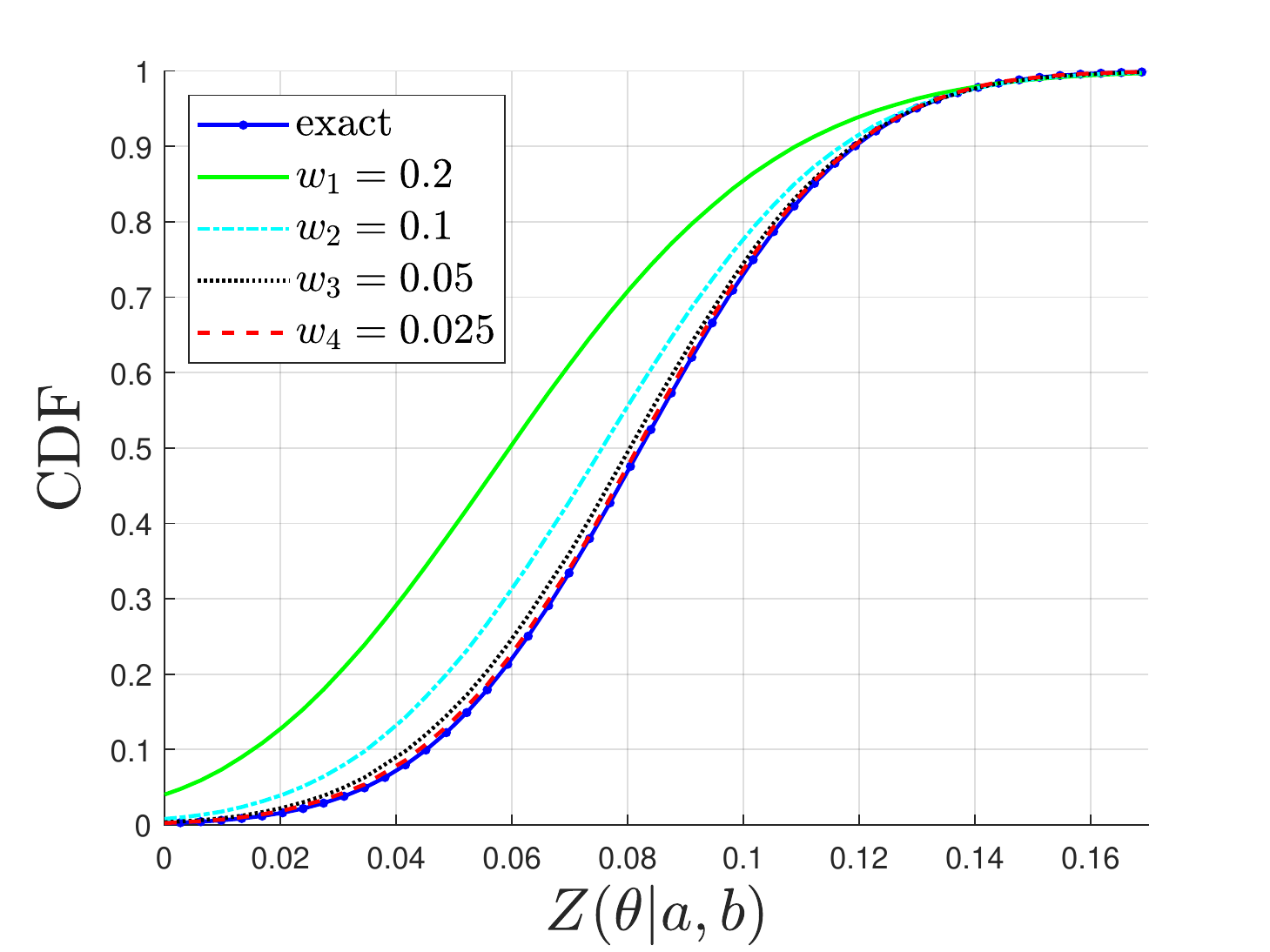} }}%
    ~
    \subfloat[\centering]{{\includegraphics[width=7.cm]{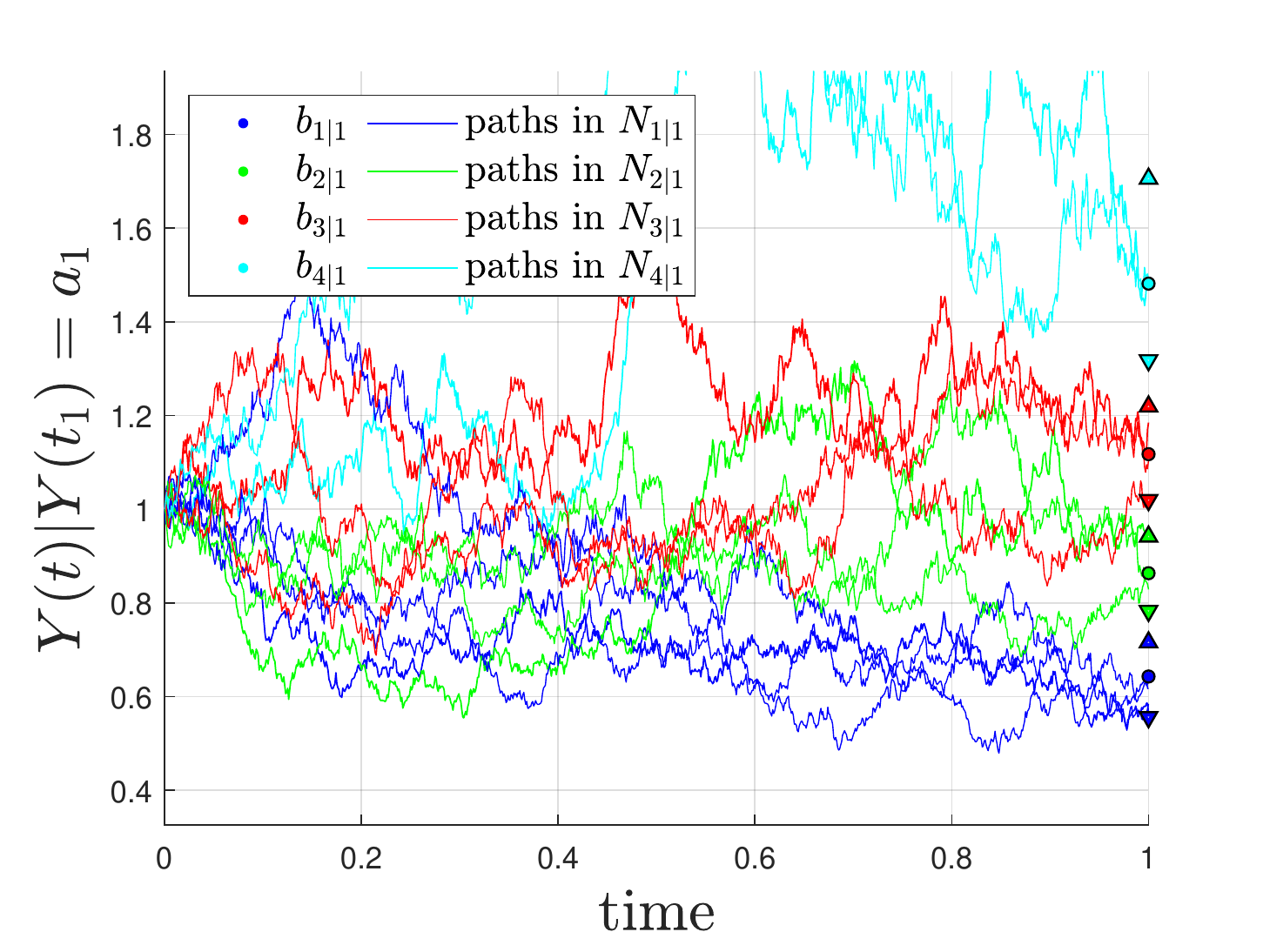} }}%
    \caption{\footnotesize Left: $Z(\theta|a_i,N(h|i))$ converges in distribution to $Z(\theta|a_i, b_{h|i})$, when the width $w$ tends to 0. Right: Different widths of the neighborhoods $N_{h|i}$ (coloured triangles) with some \emph{nearest paths}.}%
    \label{fig: error bins}%
\end{figure}

Regarding this source of error, we provide only an intuitive and informal analysis.
Using the same notation as in Remark \ref{rem: numerical computation C}, when the width of the neighborhood $N_{h|i}$ tends to zero, the distribution of $Z(\theta|a_i, N_{h|i})$ converges to that of $Z(\theta|a_i,b_{h|i})$ ($i=1,\dots,M_a$, $h=1,\dots,M_b$), as is illustrated in Figure \ref{fig: error bins}a. 

In the practical implementation, on the other hand, it is not possible to arbitrarily reduce the width of $N_{h|i}$ while keeping the total number of paths in each MC simulation unchanged.

\begin{rem}[Implementation detail]
To obtain a reliable empirical approximation of the random variable $Z(\theta|a_i,N_{h|i})$ (and consequently -- if the amplitude of $N_{h|i}$ is small enough -- also of $Z(\theta|a_i,b_{h|i})$) we need a sufficiently rich empirical population. Our choice is to fix the number of samples $2N$ a priori, which means that the width of the neighborhood $N_{h|i}$ is determined by the $N$ \emph{lower nearest} paths and the $N$ \emph{upper nearest} paths (with respect to $b_{h|i}$).
\end{rem}

In general, what has just been observed leads to different widths for different $N_{h|i}$, as shown in Figure \ref{fig: error bins}b. The neighborhoods corresponding to extreme reference (conditional) final values, namely quantiles in the tail(s) of $Y(t_1+\Dt)|Y(t_1)$, are wider and thus the approximation is less accurate. However, if the number of paths generated in MC simulations is large enough, satisfactory results can be obtained, even for extreme quantiles.

\subsubsection{Decompression error}
We now move to the error induced by calculating $\Tilde{z}_k(a,b)$ via inter/extrapolation during the decompression phase. Again, we will just provide an intuitive and non rigorous discussion.

In the proposed methodology, the interpolation rules $\psi_k$ ($k=1,\dots,M$) are constructed from Lagrange interpolating polynomials of first or second degree. Therefore, the error is a suitable function $\zeta$ of the products
\begin{equation}
    \Pi(a):=\prod_{i\in N(a)} (a-a_i), \qquad \Pi(b|a):=\prod_{h\in N(b|a)} (b - b_h(a)),
\end{equation}
with $N(a)\subset \{1,\dots,M_a\}$ (resp. $N(b|a)\subset \{1,\dots,M_b\}$) set of indices of the reference values $a_i$ (resp. $b_h(a)$) ``nearest''~\footnote{These ``nearest'' values are two or three depending on the degree of the interpolating polynomial.} to $a$ (resp. $b$), i.e.,
\begin{equation}
\label{eq: products}
    \big|z_k(a,b) - \Tilde{z}_k(a,b)\big|=\big|z_k(a,b) - \psi_k(a,b;A,B,C)\big|=:\zeta\big(\Pi(a), \Pi(b|a)\big).
\end{equation}
Note that the ``nearest'' values to $b$ are not chosen directly from the values in $B$, but are taken from the vector
\begin{equation}
    B(a):=\begin{bmatrix}
        b_1(a) & b_2(a) & \dots & b_{M_b}(a)
    \end{bmatrix},
\end{equation}
obtained from $B$ by interpolation along the ``a-direction'', i.e., these are the (approximated) references for the final value \emph{conditional} to the initial value $a$.

The function $\zeta$ -- we can imagine it as a two-dimensional generalization of Equation (\ref{eq: error for g}) -- tends to zero when $\Pi(a)$ and $\Pi(b|a)$ approach zero. We observe that when the two discretizations in $A$ and in $B$ tend to the continuous range, i.e., when $M_a$ and $M_b$ tend to infinity, the two products in Equation (\ref{eq: products}) go to zero (if $a$ and $b$ belong to the intervals covered by $A$ and $B$), and so does the error.

In the case the boundary conditions $a$ and/or $b$ do not belong to the intervals covered by $A$ and $B$ respectively, i.e., when extrapolation instead of interpolation is required, the quality of the polynomial approximation may deteriorate. However, two considerations should be made: first, the construction of the matrices $A$ and $B$ is done in such a way that the cases that need extrapolation are few compared to the total cases; moreover, by inspecting the matrix $C$ for each individual process $Y(t)$, it is possible to choose interpolating functions that produce a satisfactory result even when extrapolation is required (for an example, see Section \ref{sssec: GBM}).

\subsubsection{ANN error}
We conclude this part by studying the error due to using the artificial neural network (ANN) \cite{LiGrOo, Ya} to compute the grid of CPs $C$. 
We report below a result showing that, in principle, artificial neural networks with appropriate (simple) architectures can be used to approximate functions with an arbitrarily small error. Note, however, that despite its theoretical value, the result is of little use for implementation purposes. In fact, most of the choices made during the implementation are mainly due to empirical reasons, with particular attention to the available literature \cite{LiGrOo}.

Consider the \emph{Sobolev space} $\big(\mathcal{W}^{n,\infty}\big([0,1]^d\big), ||\cdot||_{n,d}$, with $n,d\in\mathbb{N}^*$. It can be described as the space of functions $C^{n-1}\big([0,1]^d\big)$ whose derivatives up to the $n-1^{th}$ order are Lipschitz continuous, equipped with the canonical norm $||\cdot||_{n.d}$ (typically used in the literature for Sobolev spaces). We define the unit ball $B_{n,d}:=\big\{f\in\mathcal{W}^{n,\infty}\big([0,1]^d\big): ||f||_{\mathcal{W}}\leq 1\big\}$.
The following approximation result holds (see \cite{LiGrOo, Ya}).
\begin{thm}
\label{teo: conv ANN}
For any choice of $d,n\in\mathbb{N}^*$ and $\epsilon\in(0,1)$ there exists an architecture $H(\x|\cdot)$ based on \texttt{ReLU} (Rectified Linear Unit) activation functions $\phi$, i.e. $\phi(x)=\max(x, 0)$, such that:
\begin{enumerate}
    \item $H(\x|\cdot)$ is able to approximate any function $f\in B_{d,n}$ with an error smaller than $\epsilon$, i.e., there exists a matrix of weights $\W$ such that $||f(\cdot)-H(\cdot|\W)||_{\infty}< \epsilon$;
    \item $H$ has at most $c(\ln{1/\epsilon} + 1)$ layers and at most $c\epsilon^{-d/n}(\ln{1/\epsilon}+1)$ weights and neurons, with $c=c(d,n)$ an appropriate constant function of $d$ and $n$.
\end{enumerate}
\end{thm}

\begin{rem}[Input scaling]
We emphasize that although the previous result applies to (a subclass of) functions with domain in $[0,1]^d$, this is not restrictive. In fact, in the training phase it is always possible to scale the input data such that they fall into the above domain.
\end{rem}

Theorem \ref{teo: conv ANN} provides a robust theoretical justification for the use of ANNs as regressors. The goodness of the result can also be investigated empirically, as we will show in the next section. We conclude by observing that in any case the quality of the regressor is related to the quality of the training process: all the results of convergence for ANN are based on the assumption that the training is performed correctly, and that consequently the error due to the optimization process is negligible.

\section{Numerical experiments and applications}
\label{sec: appl}
This section is dedicated to the application of the methodology described in Section \ref{sec: methodology}. Firstly, the method is applied to well-known stochastic processes. When the distribution of the conditional time-integrated process is available theoretically, this will be used as a benchmark to check the quality of the result; on the other hand, in the case the distribution is not known, we compare the result with a highly accurate empirical distribution obtained through expensive plain MC simulations. Later, these results are employed in the real application as the sampling from the Heston model and from the SABR model. Notice that since all the processes treated here are Markovian and time-invariant, from now on, the notation will be the one introduced in Remark \ref{rem: time-inv Markov processes} and not the more general used so far.

\subsection{Models}
\label{ssec: models}
Here we focus on three different stochastic processes: the Arithmetic Brownian Motion (ABM), the Geometric Brownian Motion (GBM) and the Cox-Ingersoll-Ross (CIR) process.
We start with the trivial case of ABM, which allows a theoretical comparison for the distribution of the corresponding conditional time-integrated process. Then, we move to the GBM and the CIR process, which are more attractive from the application viewpoint. However, for the GBM does not exist any theoretical comparison, while in the case of the CIR process there is, in principle, the possibility of a theoretical comparison, although it is challenging to implement and computationally expensive.

\subsubsection{Arithmetic Brownian Motion}
\label{sssec: ABM}
Let us consider the process $Y(t)$, $0\leq t\leq \Dt$, to be an ABM with model parameters $\tttheta=\{\mu,\sigma,\Dt\}$ (the model parameters are $\theta_0=\{\mu, \sigma\}$ and the ABM is a Markov process), namely its dynamics is described by
\begin{equation}
\label{eq: SDE ABM}
    \d Y(t) = \mu \dt + \sigma \d W(t).
\end{equation}

We are interested in sampling from the time-integrated bridge $Z(\theta|a,b)$ for any choice of boundary conditions $(a,b)$.
In principle, this can be done applying the whole procedure described above, even though this may be inconvenient. Since the grid $A$ is specified a priori, there is no guarantee that the initial value $a$ belongs to the range covered by the grid $A$ (hence, extrapolation is needed and the accuracy may deteriorate).

Nevertheless, the methodology gives extremely accurate results, that we briefly show together with the details of the procedure.

First we generate the training set with compression (Section \ref{sssec: compr}). We specify the range for $A$, namely $A:=[0, 1]$. We choose $M_a$, $M_b$, $M$, and so also the sizes of the three grids $A$, $B$, $C$ (Table \ref{tab: Ma, Mb, M}). The inputs in the training set are sets of model parameters $\theta=\{\mu, \sigma, \Dt\}$. The $\mu$ and $\sigma$ are randomly generated through \emph{Latine Hypercube sampling} (LHS), whereas the $\Dt$'s are taken \emph{equally-spaced} (EQ-SP) (Table \ref{tab: training set}). 
\begin{table}[h!]
\resizebox{1.\textwidth}{!}{
\parbox{.4\linewidth}{
\begin{center}
\caption{Grids dimension.}
\label{tab: Ma, Mb, M}
\vspace{-0.25cm}
\begin{tabular}{llll}
\hline\hline
\multicolumn{1}{c}{$M_a$}   & \multicolumn{1}{c|}{2}                                                          & \multicolumn{1}{c}{$A$} & \multicolumn{1}{c}{2}                                   \\
\multicolumn{1}{c}{$M_b$}   & \multicolumn{1}{c|}{2}                                                          & \multicolumn{1}{c}{$B$} & \multicolumn{1}{c}{$2\times 2$}                                   \\ \multicolumn{1}{c}{$M$}   & \multicolumn{1}{c|}{3}                                                          & \multicolumn{1}{c}{$C$} & \multicolumn{1}{c}{$2\times 2\times 3$}                  \\
\end{tabular}
\end{center}
}
\qquad
\parbox{.6\linewidth}{
\begin{center}
\caption{Training inputs.}
\label{tab: training set}
\vspace{-0.25cm}
\begin{tabular}{llll}
\hline\hline
\multicolumn{1}{c|}{$\theta$}   & \multicolumn{1}{c}{n. values}                                                   & \multicolumn{1}{c}{range} & \multicolumn{1}{c}{method}  
       \\\hline 
\multicolumn{1}{c|}{$\mu$}   & \multicolumn{1}{c}{600}                                                          & \multicolumn{1}{c}{[0.0, 0.1]} & \multicolumn{1}{c}{LHS}                                   \\ \multicolumn{1}{c|}{$\sigma$}   & \multicolumn{1}{c}{600}                                                          &
\multicolumn{1}{c}{[0.01, 0.60]}   & \multicolumn{1}{c}{LHS}   \\ \multicolumn{1}{c|}{$\Dt$} & \multicolumn{1}{c}{100} &
\multicolumn{1}{c}{[0.10, 1.09]} & \multicolumn{1}{c}{EQ-SP}\\
\end{tabular}
\end{center}
}}
\end{table}
For each pair $(\mu, \sigma)$ are run $M_a$ MC simulations to compute the corresponding grids $C$'s (one for each $t_2$).
The resulting training set $\boldsymbol{T}$ is composed by $60000$ pairs $(\theta, C)$ employed for the training of the ANN. The grid $A$ is given a priori, whereas $B$ can be computed analytically as quantiles of a suitable normal distribution obtained solving the SDE in Equation (\ref{eq: SDE ABM}), hence there is no need to train the ANN on them.

\begin{rem}[Boundaries for the grid $B$]
\label{rem: boundaries B}
In principle when $M_b$ tends to infinity the discretization along the ``$b$-direction'' converges to the continuous case. On the contrary, in the practical application the grid $C$ is computed numerically, and a big value of $M_b$ pushes the extreme values of $B$ (as $b_{1|i}$ and $b_{M_b|i}$, $i=1,\dots,M_a$) in the tails of the distribution. As a result the empirical distribution of $Z(\theta|a_i, N_{1|i})$ (resp. $Z(\theta|a_i, N_{M_b|i})$) is only a poor approximation of $Z(\theta|a_i,b_{1|i})$ (resp. $Z(\theta|a_i,b_{M_b|i})$) (see Remark \ref{rem: numerical computation C}). For this reason we prefer to put boundaries that cannot be trespassed by any value of $B$. In particular, for the ABM we set $b_{1|i}$ (resp. $b_{M_b|i}$) as the 20 \% (resp. 80 \%) quantile of the distribution of $Y(\Dt)|Y(0)=a_i$.
\end{rem}

The fully connected neural network $H$ employed in the training counts 6 layers: one input, one output and 4 hidden layers. The number of neurons in the input layer is the cardinality of $\theta$, i.e. 3, as well as the one of the output layer is the cardinality of $C$, i.e. 12. Each hidden layer has 50 neurons and we employed \texttt{Softplus} as nonlinear activation function \cite{NwIjGaMa}.

\begin{figure}[t!]%
    \centering
    \subfloat[\centering]{{\includegraphics[width=7.cm]{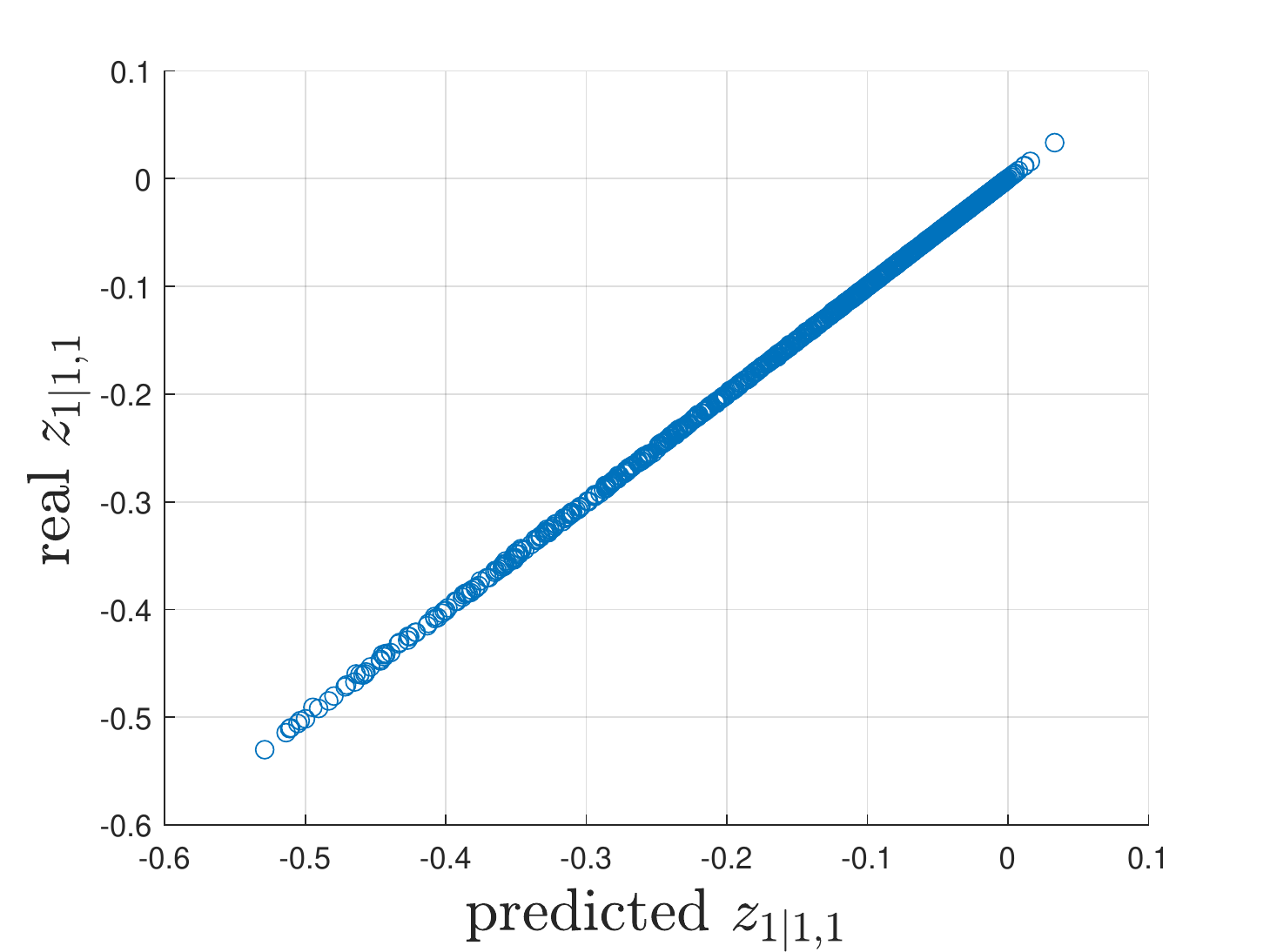} }}%
    ~
    \subfloat[\centering]{{\includegraphics[width=7.cm]{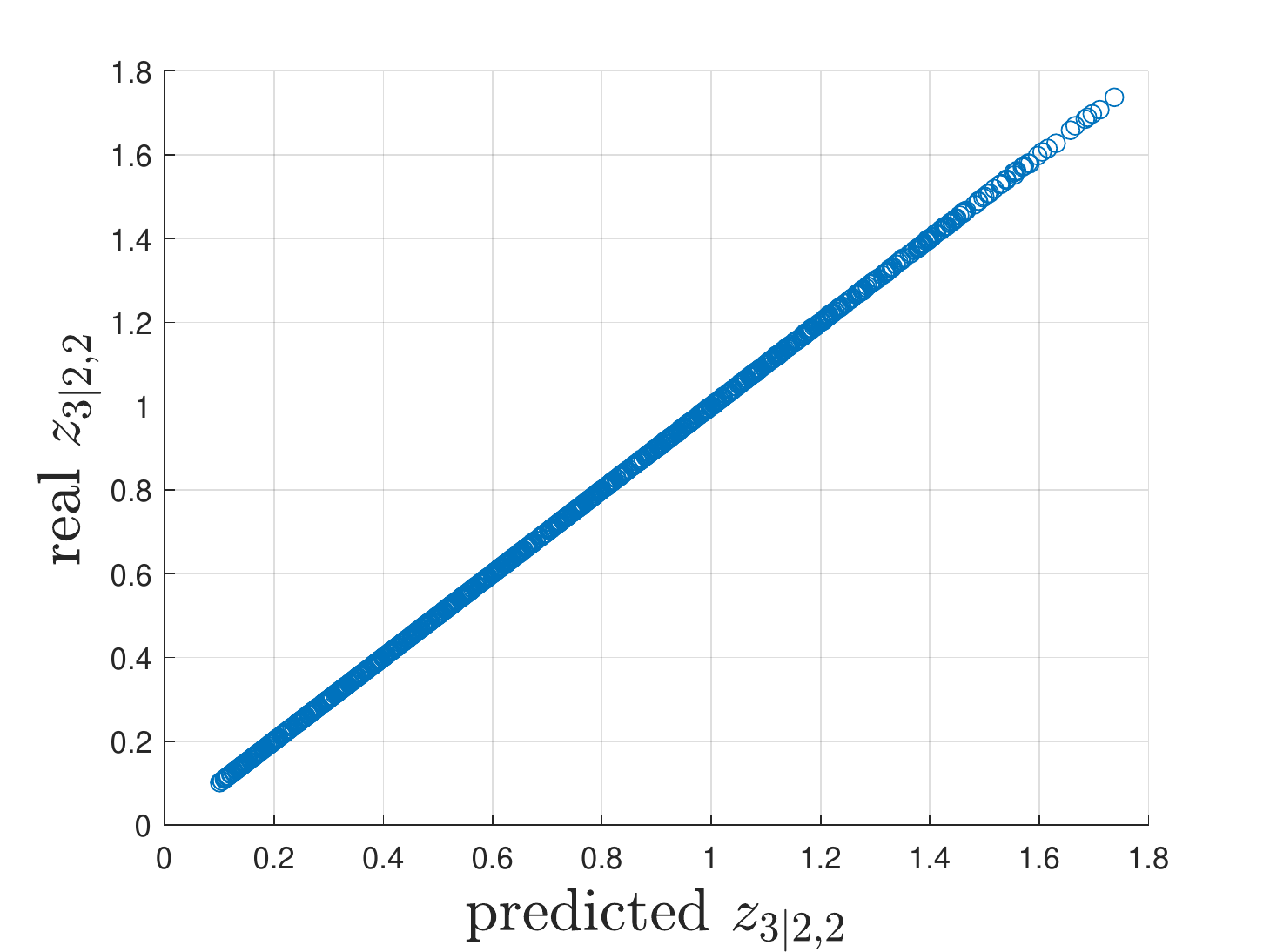} }}%
    \caption{\footnotesize Real output vs predicted output (particularly $z_{1|1,1}$ and $z_{3|2,2}$).}%
    \label{fig: accuracy ANN}%
\end{figure}

Before the training, the set $\boldsymbol T$ is splitted: the 70 \% of $\boldsymbol T$ is employed for the actual \emph{training}, the 20 \% is used for \emph{validation} -- to avoid over-fitting the training set itself -- and the remaining 10 \% for \emph{test} -- to provide a measure of accuracy of the network on unseen data. 
The parameters of $H$ are calibrated via \emph{back-propagation} using \emph{stochastic gradient descent}. The optimizer is \emph{Adam} \cite{KiBa}. The training consists in 2500 epochs. During each epoch, the whole training set (namely the $70\%$ of $\boldsymbol T$) is passed through the network in batches of size 16. The initial learning rate for the stochastic gradient descent is $10^{-3}$ and it halves every 500 epochs. The trained ANN maps the input parameters $\theta$ into the corresponding grid $C$. The ranges of the output collocation points are reported in Table \ref{tab: outputs range}. As we can see in Figure \ref{fig: accuracy ANN}, the ANN reaches a high level of accuracy on the test set. This concludes the off-line phase.

\begin{table}[b]
\begin{center}
\caption{Ranges entries matrix C.}
\label{tab: outputs range}
\vspace{-0.25cm}
\resizebox{0.8\textwidth}{!}{
\begin{tabular}{lllll}
\hline\hline
\multicolumn{1}{c|}{$a_i$}&
\multicolumn{2}{c}{$a_1$} & 
\multicolumn{2}{c}{$a_2$}
       \\
\multicolumn{1}{c|}{$b_{h|i}$}&
\multicolumn{1}{c}{$b_{1|1}$} & \multicolumn{1}{c}{$b_{2|1}$}  & 
\multicolumn{1}{c}{$b_{1|2}$} & \multicolumn{1}{c}{$b_{2|2}$}  \\\hline
\multicolumn{1}{c|}{$z_{1|i,h}$}&
\multicolumn{1}{c}{[-0.631, 0.049]} &
\multicolumn{1}{c}{[-0.055, 0.058]} &
\multicolumn{1}{c}{[0.083, 1.139]} &
\multicolumn{1}{c}{[0.099, 1.149]} \\
\multicolumn{1}{c|}{$z_{2|i,h}$}&
\multicolumn{1}{c}{[-0.288, 0.055]} &
\multicolumn{1}{c}{[0.000, 0.347]} &
\multicolumn{1}{c}{[0.092, 1.145]} &
\multicolumn{1}{c}{[0.100, 1.437]} \\
\multicolumn{1}{c|}{$z_{3|i,h}$}&
\multicolumn{1}{c}{[0.000, 0.115]} &
\multicolumn{1}{c}{[0.0001, 0.690]} &
\multicolumn{1}{c}{[0.099, 1.204]} &
\multicolumn{1}{c}{[0.1001, 1.780]} \\
\end{tabular}}
\end{center}
\end{table}

At this point, the sampling can be performed, taking advantage of the fast on-line part. In application, we usually have one set of parameters $\theta$, and we want to sample from different conditional time-integrated bridges simultaneously because the boundary conditions for each sample might be different. Therefore, in the more general case, we have an array of initial and final values. Notice that the length of the two arrays of boundary conditions has to be the same, and it corresponds to the overall number of samples. The procedure is exactly the one described in the algorithm at the end of Section \ref{sec: methodology}. The only note concerns the interpolation rules $\psi_k(\cdot,\cdot;A,B,C)$'s.
\begin{rem}[Interpolation criterion for conditional time-integrated ABM]
Inspecting the matrix $C$ linear interpolation (LI) is the best choice along both the ``$a$-direction'' and the ``$b$-direction''. Particularly, linear extrapolation (LE) gives accurate results also when the boundary conditions are outside the ranges covered by $A$ and/or $B$, as shown in Table \ref{tab: result procedure}. Moreover, when inter/extrapolation is performed, small $M_a$ and $M_b$ are desirable to avoid the propagation and the increase of the natural errors due to the numerical construction of the grid $C$ in the compression phase (see Remark \ref{rem: numerical computation C}).
\end{rem}
\begin{table}[b]
\begin{center}
\caption{Application details and results.}
\label{tab: result procedure}
\resizebox{0.85\textwidth}{!}{
\begin{tabular}{llllllllll}
\hline\hline
\multicolumn{2}{c}{\multirow{2}{*}{Parameters}} & \multicolumn{1}{c}{}&
\multicolumn{7}{c}{\multirow{2}{*}{Error $\epsilon$ \quad($\times\: 10^{-3}$)}}\\
\multicolumn{2}{c}{\multirow{2}{*}{}} & \multicolumn{1}{c}{}&
\multicolumn{7}{c}{}
       \\\cline{1-2}\cline{4-10}
\multicolumn{1}{c|}{$\mu$} & 
\multicolumn{1}{c}{0.04} &
\multicolumn{1}{c}{} &
\multicolumn{2}{c|}{\multirow{2}{*}{$(a_n, b_m)$}} &
\multicolumn{1}{c}{Q 35\%}&
\multicolumn{1}{c}{Q 55\%}&
\multicolumn{1}{c}{Q 75\%}&
\multicolumn{1}{c}{Q 95\%}&
\multicolumn{1}{c}{}\\
\multicolumn{1}{c|}{$\sigma$} & 
\multicolumn{1}{c}{0.3} & \multicolumn{1}{c}{} &
\multicolumn{1}{c}{} &
\multicolumn{1}{c|}{} &
\multicolumn{1}{c}{$b_1$} &
\multicolumn{1}{c}{$b_2$} &
\multicolumn{1}{c}{$b_3$} &
\multicolumn{1}{c}{$b_4$}&
\multicolumn{1}{c}{} 
       \\\cline{4-10}  
            \multicolumn{1}{c|}{$\Dt$} & 
\multicolumn{1}{c}{1.0} &
\multicolumn{1}{c}{} &
\multicolumn{1}{c}{$a_1$} &
\multicolumn{1}{c|}{0.8} &
\multicolumn{1}{c}{2.5} &
\multicolumn{1}{c}{3.8} &
\multicolumn{1}{c}{3.5} &
\multicolumn{1}{c|}{3.8} &
\multicolumn{1}{c}{LI} 
       \\
        \multicolumn{2}{c}{\multirow{2}{*}{Simulation}} &
\multicolumn{1}{c}{} &
\multicolumn{1}{c}{$a_2$} &
\multicolumn{1}{c|}{0.9} &
\multicolumn{1}{c}{2.4} &
\multicolumn{1}{c}{4.8} &
\multicolumn{1}{c}{1.8} &
\multicolumn{1}{c|}{3.2} &
\multicolumn{1}{c}{LI} 
       \\
       \multicolumn{2}{c}{\multirow{2}{*}{}} &
      \multicolumn{1}{c}{} &   
\multicolumn{1}{c}{$a_3$} &
\multicolumn{1}{c|}{1.0} &
\multicolumn{1}{c}{2.2} &
\multicolumn{1}{c}{3.7} &
\multicolumn{1}{c}{3.3} &
\multicolumn{1}{c|}{3.0}&
\multicolumn{1}{c}{LI} 
       \\\cline{1-2}
\multicolumn{1}{c|}{samples} & \multicolumn{1}{c}{$10^5$}&
\multicolumn{1}{c}{} &
\multicolumn{1}{c}{$a_4$} &
\multicolumn{1}{c|}{1.1} &
\multicolumn{1}{c}{2.7} &
\multicolumn{1}{c}{2.8} &
\multicolumn{1}{c}{3.0} &
\multicolumn{1}{c|}{1.7}&
\multicolumn{1}{c}{LE} \\\cline{4-10}
       
\multicolumn{1}{c|}{time (msec)} &
       \multicolumn{1}{c}{120} & 
\multicolumn{1}{c}{} &
\multicolumn{1}{c}{} &
\multicolumn{1}{c|}{}  &
\multicolumn{1}{c}{LI} &
\multicolumn{1}{c}{LI} &
\multicolumn{1}{c}{LI} &
\multicolumn{1}{c|}{LE}
&
\multicolumn{1}{c}{rule}
\end{tabular}}
\end{center}
\end{table}

\begin{figure}[t]%
    \centering
    \subfloat[\centering]{{\includegraphics[width=6.4cm]{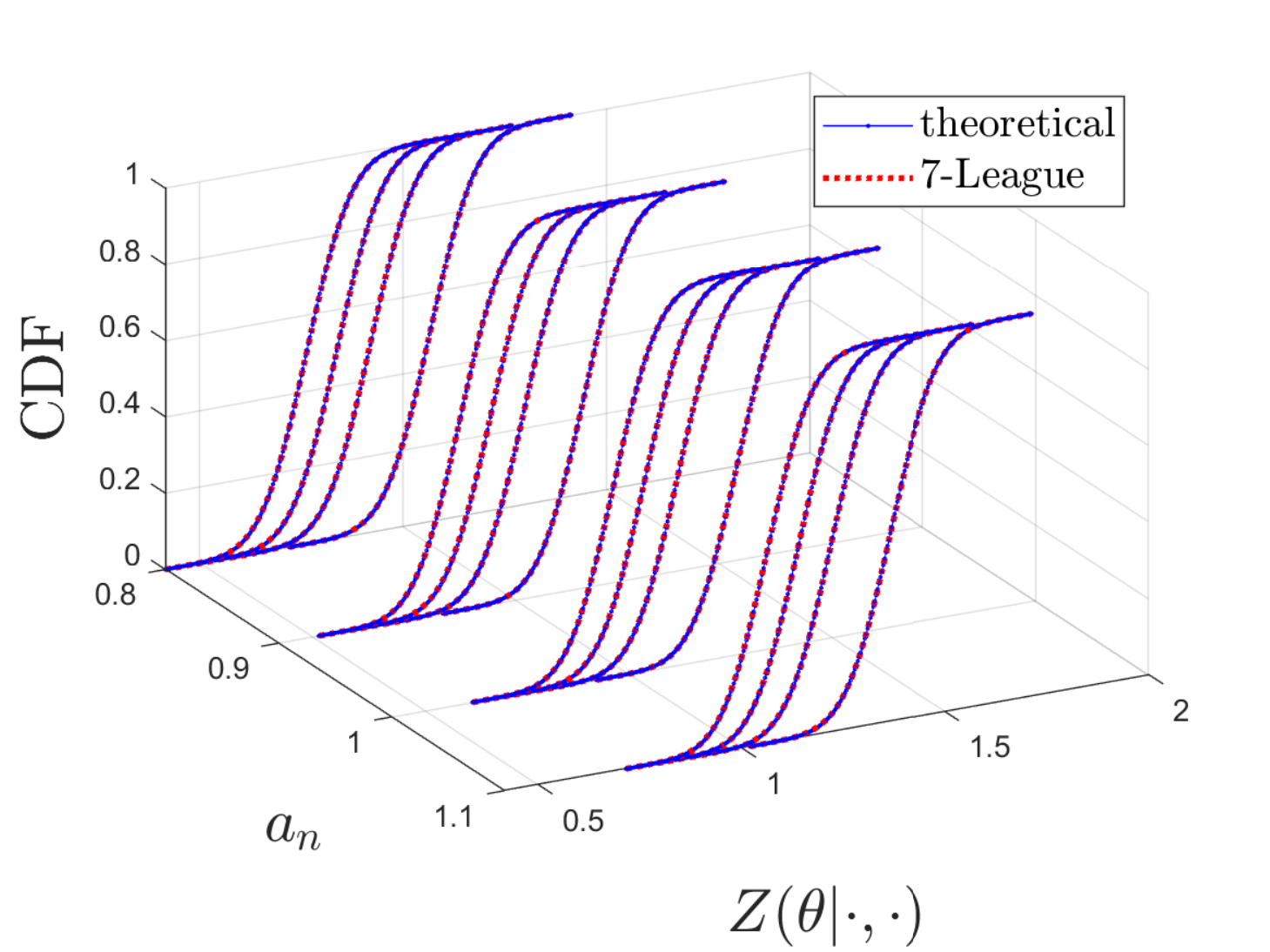} }}%
    ~
    \subfloat[\centering]{{\includegraphics[width=6.1cm]{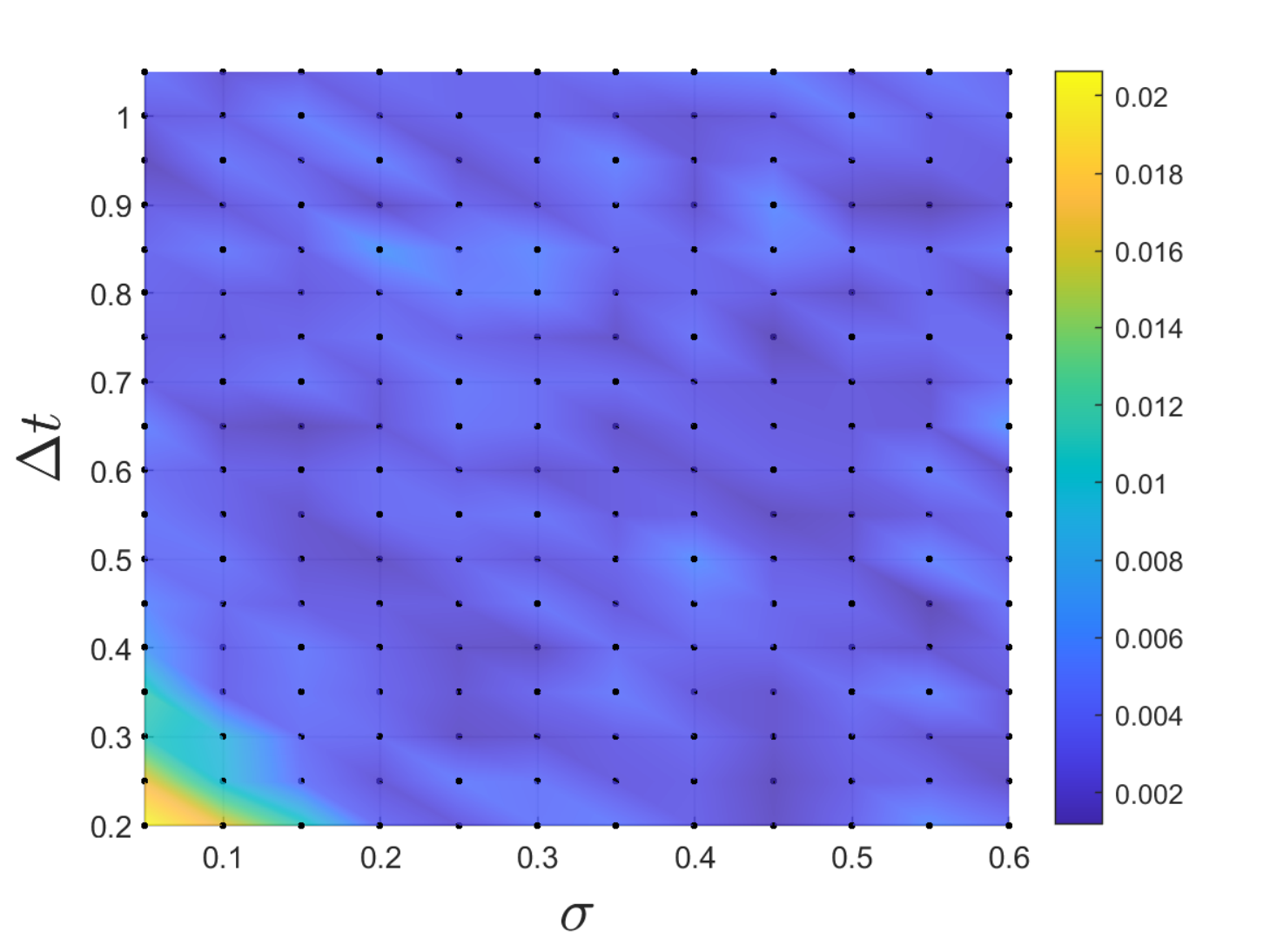} }}%
    \caption{\footnotesize Left: Comparison between empirical and theoretical CDFs varying the boundary conditions $(a,b)$. Right: Heat-map of the error $\epsilon$ varying the model parameters $\sigma$ and $\Dt$ (with a fixed $\mu=0.04$). The boundary conditions are $a=0.5$ and $b$ is taken as the 95\% quantiles.}%
    \label{fig: results ABM}%
\end{figure}

The results of the application of the methodology to the time-integrated ABM are given in Table \ref{tab: result procedure} and in Figure \ref{fig: results ABM}. In Figure \ref{fig: results ABM}a we show the comparison between the empirical CDFs obtained as result of the methodology and the exact theoretical CDFs (available in the case of ABM). The different CDFs are obtained varying the pair of boundary conditions $(a,b)$, while keeping fixed the set of model parameters (and $\Dt$) $\theta$. In particular, the initial values for $a$ are equally-spaced on a given range, whereas the final values are computed as quantiles of the conditional distributions $Y(\Dt)|Y(0)=a_n$, $n=1,2,3,4$ (see Table \ref{tab: result procedure}). We can notice that the empirical CDFs (red dashed lines) are almost indistinguishable from the theoretical ones (blue pointed lines) even for extreme values of the boundary condition $b$ (quantile 95\%). The goodness of the results is confirmed in Table \ref{tab: result procedure} where the errors are reported together with the interpolation rule employed in the decompression, along both the ``$a$-direction'' and the ``$b$-direction''. The error is defined as the $L^{\infty}$-distance between the two CDFs, namely
\begin{equation}
\label{eq: error definition}
    \epsilon = \underset{x}{\text{sup}} |F_{M}(x) - F_{B}(x)|,
\end{equation}
where $F_{M}$ is the empirical CDF obtained by the application of the methodology and $F_{B}$ is a benchmark CDF (in this case it is the theoretical one).
There is no significant deterioration in the result when extrapolation is employed. In Table \ref{tab: result procedure} are available also the details about the procedure, as the number of samples computed for each pair of boundary conditions and the corresponding computational time. On the other hand, Figure \ref{fig: results ABM}b shows a heat-map of the errors when the model parameters (and $\Dt$) vary. Particularly, the errors reported are referred to the distribution $Z(\theta|0,b(\theta))$, with $b(\theta)$ the 95\% quantile of the process at final time $\Dt$. The result is extremely satisfactory, indeed the error is most of the time below 0.01 (even for extreme values for the final value $b$, as the 95\% quantile!). The only exceptions are when either $\sigma$ or $\Dt$ are small. Such a lack of accuracy is mainly due to two reasons: first, the quality of the training set deteriorates for small values of $\sigma$ and $\Dt$ (in principle this issue could be solved increasing the number of paths in each MC simulation during the construction of the training set); then, the accuracy of the ANN tends to decrease close to the boundaries of the training domain. Even though the application of the proposed procedure to the conditional time-integrated ABM is useless from a practical point of view (since the analytic distribution is known), it provides a bright evidence that the methodology can produce extremely accurate samples with a low computational time, with respect to plain MC simulations. In the following other models are presented, which are more interesting from the application viewpoint.

\subsubsection{Geometric Brownian Motion}
\label{sssec: GBM}

Let us assume the process $Y(t)$ to be a GBM with model parameters $\tttheta=\{\mu,\sigma,\Dt\}$, i.e. driven by the dynamics
\begin{equation}
    \d Y(t) = \mu Y(t)\dt + \sigma Y(t) \d W(t).
\end{equation}
As for the previous ABM example, the methodology could be directly applied, but this is not the most efficient way to tackle the problem of sampling from conditional time-integrated GBM.
Indeed, we can fully avoid the problem of specifying a priori the grid $A$ exploiting the following \emph{scaling} on the original process $Y(t)$.

\begin{rem}[GBM scaling]
Let we consider the GBM, $Y(t)$, with $Y(0)=a>0$. The original process $Y(t)$ can be rewritten in terms of a ``standard'' one $\hat Y(t)$ starting at 1, namely
\begin{equation}
\label{eq: scale GBM}
    Y(t)=a \hat Y(t).
\end{equation}
From {It\^o's lemma}, the dynamics of the process $\hat Y(t)$ is exactly the same as the one of the original process $Y(t)$.
Given the final condition $b$, Equation (\ref{eq: scale GBM}) entails the following relationship between the corresponding time-integrated bridges
\begin{equation}
\label{eq: scale integral GBM}
\begin{aligned}
    Z(\theta|a,b)&=\int_{0}^{\Dt} Y(t) \d t \Big| Y(0)=a, Y(\Dt)=b\\
    &= a\int_{0}^{\Dt} \hat Y(t) \d t \Big| \hat Y(0)=1, \hat Y(\Dt)=b/a\\
    &=a Z(\theta|1,b/a).
\end{aligned}
\end{equation}
\end{rem}
Equation (\ref{eq: scale integral GBM}) implies that for any positive $a$ the samples from any conditional integral $Z(\theta|a,b)$ are easily obtainable from the ones of $Z(\theta|1,b/a)$.
As a consequence, the grid $A$ for the initial value collapses to the singleton 1 and then the dimensions of $B$ and $C$ decrease of one, speeding-up both the compression and the decompression phases.

The procedure is applied as described in Section \ref{sec: methodology}.
The training set is built similarly as shown in Section \ref{sssec: ABM} and the details are specified in Table \ref{tab: Ma, Mb, M GBM} and Table \ref{tab: training set GBM}.
\begin{table}[H]
\resizebox{1.\textwidth}{!}{\parbox{.4\linewidth}{
\begin{center}
\caption{Grids dimension.}
\label{tab: Ma, Mb, M GBM}
\vspace{-0.25cm}
\begin{tabular}{llll}
\hline\hline
\multicolumn{1}{c}{$M_a$}   & \multicolumn{1}{c|}{1}                                                          & \multicolumn{1}{c}{$A$} & \multicolumn{1}{c}{1}                                   \\
\multicolumn{1}{c}{$M_b$}   & \multicolumn{1}{c|}{6}                                                          & \multicolumn{1}{c}{$B$} & \multicolumn{1}{c}{$1\times 6$}                                   \\ \multicolumn{1}{c}{$M$}   & \multicolumn{1}{c|}{4}                                                          & \multicolumn{1}{c}{$C$} & \multicolumn{1}{c}{$1\times 6\times 4$}                  \\
\end{tabular}
\end{center}
}
\qquad
\parbox{.6\linewidth}{
\begin{center}
\caption{Training inputs.}
\label{tab: training set GBM}
\vspace{-0.25cm}
\begin{tabular}{llll}
\hline\hline
\multicolumn{1}{c|}{$\theta$}   & \multicolumn{1}{c}{n. values}                                                   & \multicolumn{1}{c}{range} & \multicolumn{1}{c}{method}  
       \\\hline 
\multicolumn{1}{c|}{$\mu$}   & \multicolumn{1}{c}{600}                                                          & \multicolumn{1}{c}{[0.0, 0.1]} & \multicolumn{1}{c}{LHS}                                   \\ \multicolumn{1}{c|}{$\sigma$}   & \multicolumn{1}{c}{600}                                                          &
\multicolumn{1}{c}{[0.05, 0.60]}   & \multicolumn{1}{c}{LHS}   \\ \multicolumn{1}{c|}{$\Dt$} & \multicolumn{1}{c}{100} &
\multicolumn{1}{c}{[0.10, 1.09]} & \multicolumn{1}{c}{EQ-SP}\\
\end{tabular}
\end{center}
}}
\end{table}
\vspace{-0.5cm}
The ANN employed has the same architecture as the one used for the conditional time-integrated ABM (except for the output layer, which has 24 components instead of 12), and the training is performed as shown for the conditional time-integrated ABM (number of epochs, batch size, learning rates).

\begin{figure}[b!]%
    \centering
    \subfloat[\centering]{{\includegraphics[width=7.cm]{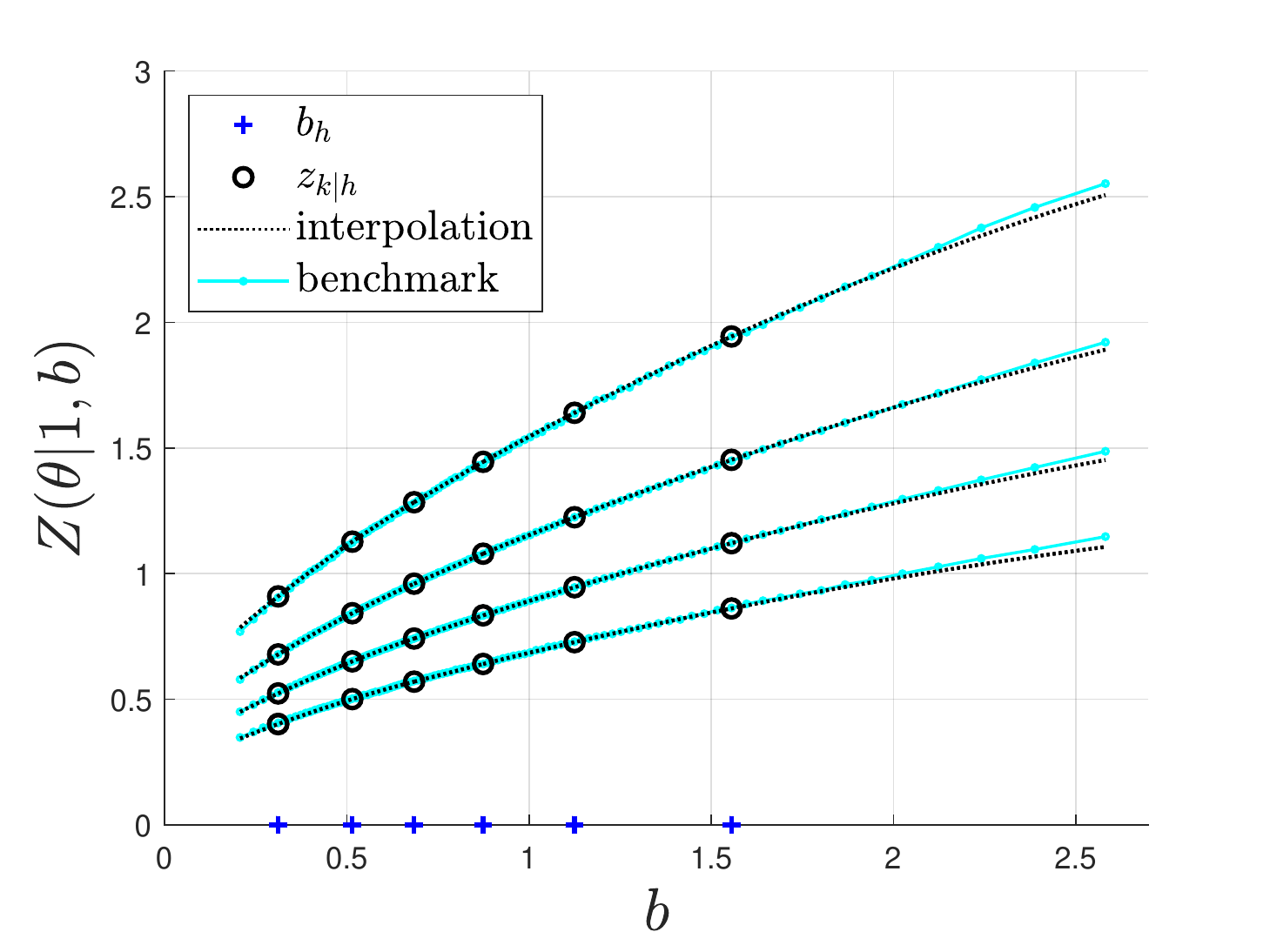} }}%
    ~
    \subfloat[\centering]{{\includegraphics[width=7.cm]{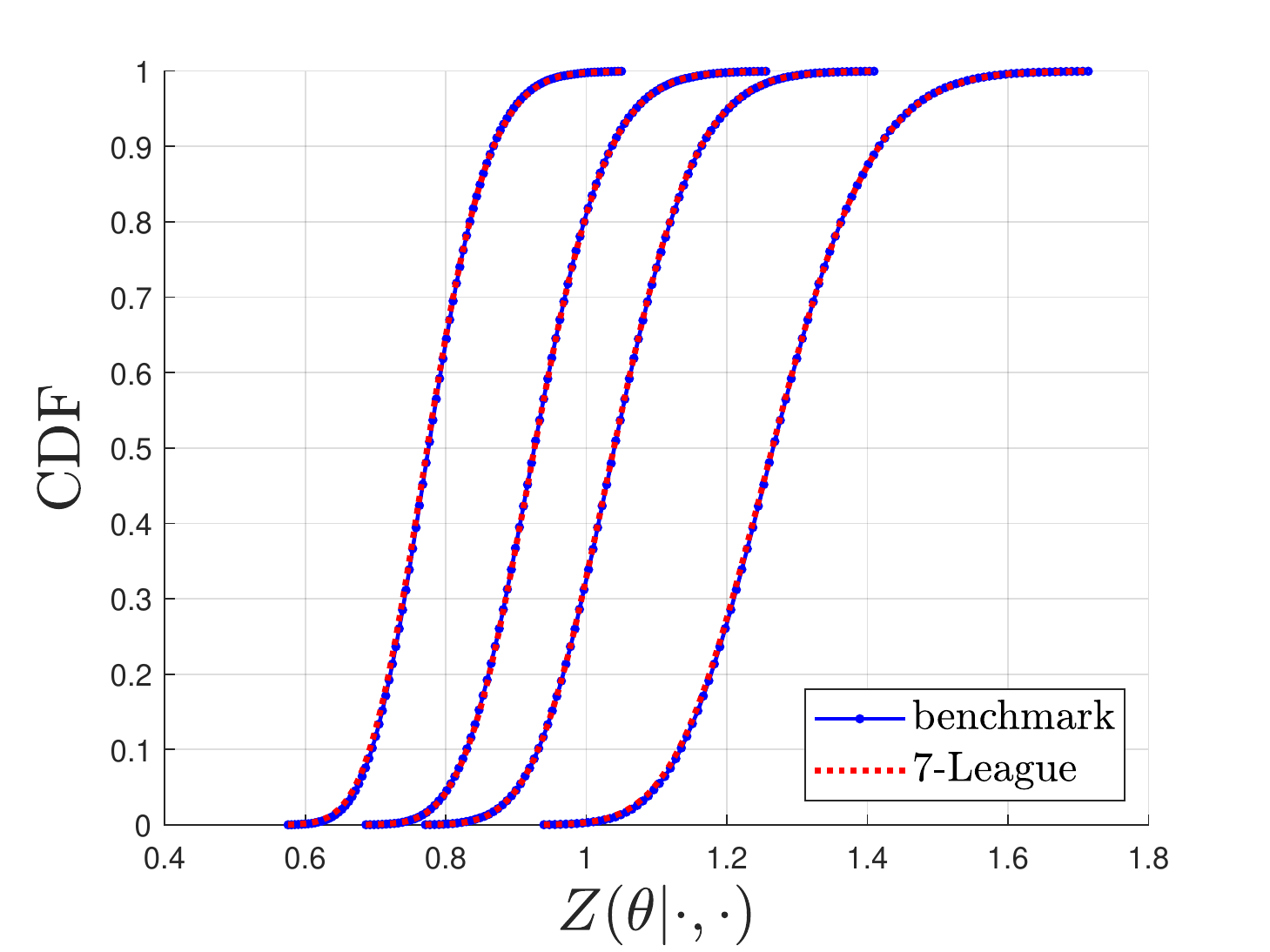} }}%
    \caption{Left: Piecewise quadratic inter/extrapolation (linear between $b_3$ and $b_4$). Right: CDFs comparison, with boundary conditions $a=1$ and $b$'s are (from the left) the 5\%, 35\%, 65\%, 95\% quantiles of $Y(\Dt)|Y(0)=1$.}%
    \label{fig: results GBM}%
\end{figure}

The most significant differences concerning the previous example concern the boundaries for the grid $B$ and the choice of the interpolation rule in the decompression phase. We decided to employ as the boundary for the grid $B$ the 5\% and the 85\% quantiles. This choice is driven by that log-normal distributions have only one infinite end-point (at $+\infty$), whereas the other is finite (at $0$). As a consequence, approaching the left end-point 0 introduces a smaller error than approaching the (infinite) right end-point (see Remarks \ref{rem: numerical computation C} and \ref{rem: boundaries B}). Regarding the interpolation rules, inspecting the grid $C$ we decided to employ a higher-order polynomial inter/extrapolation. Particularly, because of the concavity of the grid $C$ (with respect to the values in $B$), a second-degree polynomial properly fits the shape of the collocation points, increasing the accuracy of the extrapolation on extreme values significantly (see Figure \ref{fig: results GBM}a). Eventually, in Figure \ref{fig: results GBM}b are shown the empirical CDFs got applying the procedure, compared with accurate benchmarks computed via expensive MC simulation. The empirical CDFs achieve a high level of accuracy, even for extreme final values $b$ (as the quantiles 5\% and 95\%). Moreover, due to the simpler structure of the grids, the computational time for a sample of size $10^5$ is about 70 milliseconds, roughly half of the time required for the conditional time-integrated ABM.

\subsubsection{Cox-Ingersoll-Ross process}
\label{sssec: CIR}
The last model we present here is the time-integrated Cox-Ingersoll-Ross (CIR) process.
The dynamics of the process $Y(t)$, $\tttheta=\{\kappa, \overline{Y}, \gamma,\Dt\}$, is given in terms of the SDE
\begin{equation}
    \d Y(t) = \kappa (\overline{Y}- Y(t))\d t + \gamma \sqrt{Y(t)} \d W(t),
\end{equation}
where the constants $\kappa, \overline{Y}, \gamma>0$ are respectively the mean-reverting speed, the long-term mean value and the volatility of the process.

Unfortunately, no scaling is available to simplify the overall procedure, hence we have to specify a priori the grid $A$. Based on the fact that the CIR process is a \emph{mean-reverting} process, we specify the grid $A$ as a function of the long-term mean value, particularly the range covered by $A$ is $\big[0.1\cdot\overline{Y}, 2.5\cdot\overline{Y}\big]$. The off-line stage is implemented as shown previously, and the details are reported in Table \ref{tab: Ma, Mb, M CIR} and in Table \ref{tab: training set CIR}.
\begin{table}[h]
\resizebox{1.\textwidth}{!}{
\parbox{.4\linewidth}{
\begin{center}
\caption{Grids dimension.}
\label{tab: Ma, Mb, M CIR}
\vspace{-0.25cm}
\begin{tabular}{llll}
\hline\hline
\multicolumn{1}{c}{$M_a$}   & \multicolumn{1}{c|}{6}                                                          & \multicolumn{1}{c}{$A$} & \multicolumn{1}{c}{6}                                   \\
\multicolumn{1}{c}{$M_b$}   & \multicolumn{1}{c|}{6}                                                          & \multicolumn{1}{c}{$B$} & \multicolumn{1}{c}{$6\times 6$}                                   \\ \multicolumn{1}{c}{$M$}   & \multicolumn{1}{c|}{4}                                                          & \multicolumn{1}{c}{$C$} & \multicolumn{1}{c}{$6\times 6\times 4$}                  \\
\end{tabular}
\end{center}
}
\qquad
\parbox{.6\linewidth}{
\begin{center}
\caption{Training inputs.}
\label{tab: training set CIR}
\vspace{-0.25cm}
\begin{tabular}{llll}
\hline\hline
\multicolumn{1}{c|}{$\theta$}   & \multicolumn{1}{c}{n. values}                                                   & \multicolumn{1}{c}{range} & \multicolumn{1}{c}{method}  
       \\\hline 
\multicolumn{1}{c|}{$\kappa$}   & \multicolumn{1}{c}{500}                                                          & \multicolumn{1}{c}{[0.5, 1.5]} & \multicolumn{1}{c}{LHS}                                   \\ \multicolumn{1}{c|}{$\overline{Y}$}   & \multicolumn{1}{c}{500}                                                          & \multicolumn{1}{c}{[0.01, 0.50]} & \multicolumn{1}{c}{LHS}                                   \\ \multicolumn{1}{c|}{$\gamma$}   & \multicolumn{1}{c}{500}                                                          &
\multicolumn{1}{c}{[0.1, 0.5]}   & \multicolumn{1}{c}{LHS}   \\ \multicolumn{1}{c|}{$\Dt$} & \multicolumn{1}{c}{100} &
\multicolumn{1}{c}{[0.10, 1.09]} & \multicolumn{1}{c}{EQ-SP}\\
\end{tabular}
\end{center}
}}
\end{table}
The architecture of the ANN is the same as in the previous two examples, with appropriate modifications to the input and output layers. The training hyper-parameters are the same as well. The boundaries for $B$ are specified as the 10\% and 90\% quantiles, and the inter/extrapolation is linear along both the ``$a$-direction'' and the ``$b$-direction''. Even in this case, the resulting methodology is satisfactory. A measure of the goodness of the method is provided in the next section, where the methodology is compared with a full MC simulation for the log price of stock under the Heston model assumptions.

\begin{rem}[Feller condition and small initial values]
The CIR process is non-negative, provides a non-negative initial condition. Nonetheless, if the Feller condition, i.e., $2\kappa\overline{Y}>\gamma^2$, does not hold, the process may hit 0 with a strictly positive probability. Often CIR processes for which the Feller condition is not satisfied are challenging to deal with from a numerical perspective due to the presence of the atom in 0. On the other hand, the integral over time does not suffer the same problem since it is strictly positive, given a positive initial condition. Nonetheless, extremely small initial values $Y(0)$ may entail the failure of the methodology when the Feller condition does not hold.
\end{rem}

\subsection{Financial applications}
\label{ssec: fin applic}
This part of the article is dedicated to the application of the methodology to real financial situations. First, we show an alternative to the usual plain MC simulation to sample an asset's log-price under the Heston model assumptions. Eventually, we quickly show how the methodology could be accurate and effective during the sampling from the SABR model.

\subsubsection{Sampling from the Heston model}
\label{sssec: Heston}
\begin{table}[t]
\begin{center}
\caption{Heston parameters sets, with interpolation rule for the initial value $v(0)$ (L: linear, I/E: inter/extrapolation).}\vspace{-0.2cm}
\label{tab: Heston params}
\resizebox{0.6\textwidth}{!}{
\begin{tabular}{lllllll}
\hline\hline
\multicolumn{1}{c}{\multirow{2}{*}{}} & \multicolumn{1}{c}{\multirow{2}{*}{Set I}}&
\multicolumn{1}{c}{\multirow{2}{*}{Set II}}&
\multicolumn{1}{c}{\multirow{2}{*}{Set III}}&
\multicolumn{1}{c}{\multirow{2}{*}{Set IV}}&
\multicolumn{1}{c}{\multirow{2}{*}{Set V}}&
\multicolumn{1}{c}{\multirow{2}{*}{Set VI}}\\
\multicolumn{1}{c}{}&
\multicolumn{1}{c}{}&
\multicolumn{1}{c}{}&
\multicolumn{1}{c}{}&
\multicolumn{1}{c}{}&
\multicolumn{1}{c}{}\\\hline
\multicolumn{1}{c|}{$\Dt$}&
\multicolumn{1}{c}{0.5}&
\multicolumn{1}{c}{0.5}&
\multicolumn{1}{c}{1.0}&
\multicolumn{1}{c}{1.0}&
\multicolumn{1}{c}{1.0}&
\multicolumn{1}{c}{1.0}\\
\multicolumn{1}{c|}{$\kappa$}&
\multicolumn{1}{c}{1.0}&
\multicolumn{1}{c}{0.6}&
\multicolumn{1}{c}{1.0}&
\multicolumn{1}{c}{1.0}&
\multicolumn{1}{c}{0.6}&
\multicolumn{1}{c}{0.6}\\
\multicolumn{1}{c|}{$\overline{v}$}&
\multicolumn{1}{c}{0.4}&
\multicolumn{1}{c}{0.05}&
\multicolumn{1}{c}{0.4}&
\multicolumn{1}{c}{0.4}&
\multicolumn{1}{c}{0.05}&
\multicolumn{1}{c}{0.05}\\
\multicolumn{1}{c|}{$\gamma$}&
\multicolumn{1}{c}{0.2}&
\multicolumn{1}{c}{0.4}&
\multicolumn{1}{c}{0.2}&
\multicolumn{1}{c}{0.2}&
\multicolumn{1}{c}{0.4}&
\multicolumn{1}{c}{0.4}\\\hline
\multicolumn{1}{c|}{\multirow{2}{*}{$v(0)$}}&
\multicolumn{1}{c}{$\overline{v}$}&
\multicolumn{1}{c}{$\overline{v}$}&
\multicolumn{1}{c}{$\overline{v}$}&
\multicolumn{1}{c}{$0.5\cdot \overline{v}$}&
\multicolumn{1}{c}{$\overline{v}$}&
\multicolumn{1}{c}{$3\cdot \overline{v}$}\\
\multicolumn{1}{c|}{}&
\multicolumn{1}{c}{LI}&
\multicolumn{1}{c}{LI}&
\multicolumn{1}{c}{LI}&
\multicolumn{1}{c}{LI}&
\multicolumn{1}{c}{LI}&
\multicolumn{1}{c}{LE}\\
\end{tabular}}
\end{center}\vspace{-0.5cm}
\end{table}

In Section \ref{sssec: intro Heston} we presented the theory concerning the Heston Stochastic Volatility model. Here, we show the high-quality results we can get in this framework by applying the methodology proposed. This will provide an indirect measure of the goodness of the methodology.

We recall that several parameters characterize the dynamic of the Heston model: the three parameters of the CIR process $\kappa$, $\overline{v}$, $\gamma$, the interest rate $r$, the correlation coefficient $\rho$ between the two Brownian Motions, and the initial values for the variance process $v(0)$ and for the log-price process $X(0)$. We specified $r=0.01$, $\rho=-0.5$ and $X(0)=0$, whereas for the remaining parameters and for the maturity $\Dt$ we tested different choices as given in Table \ref{tab: Heston params}.

We compared the empirical CDFs obtained from the methodology application (7L) with the empirical CDFs computed through the Euler-Maruyama Monte Carlo scheme (MC) with constant time-step $\delta=0.01$.
As reported in Table \ref{tab: Heston results}, the error between the CDFs, that is, the $L^{\infty}$-distance between the curves (see Equation \ref{eq: error definition}), is very low, and on the other hand, the computational time is reduced significantly. In Figure \ref{fig: applications}a is provided with an illustration of the comparison between the empirical CDF obtained from the plain MC simulation and the one obtained using the methodology proposed.

\begin{table}[h]
\begin{center}
\caption{Heston simulation results. Comparison between Monte Carlo simulation (MC) and the methodology proposed (7L).}\vspace{-0.2cm}
\label{tab: Heston results}
\resizebox{0.8\textwidth}{!}{
\begin{tabular}{llllllll}
\hline\hline
\multicolumn{1}{c}{\multirow{2}{*}{}} & \multicolumn{1}{c}{\multirow{2}{*}{Set I}}&
\multicolumn{1}{c}{\multirow{2}{*}{Set II}}&
\multicolumn{1}{c}{\multirow{2}{*}{Set III}}&
\multicolumn{1}{c}{\multirow{2}{*}{Set IV}}&
\multicolumn{1}{c}{\multirow{2}{*}{Set V}}&
\multicolumn{1}{c}{\multirow{2}{*}{Set VI}}\\
\multicolumn{7}{c}{}&
\multicolumn{1}{c}{}\\\hline
\multicolumn{1}{c|}{error ($\times\:10^{-3}$)}&
\multicolumn{1}{c}{3.9}&
\multicolumn{1}{c}{3.6}&
\multicolumn{1}{c}{5.4}&
\multicolumn{1}{c}{3.5}&
\multicolumn{1}{c}{3.1}&
\multicolumn{1}{c|}{5.0}&
\multicolumn{1}{c}{samples}\\
\multicolumn{1}{c|}{time MC (sec)}&
\multicolumn{1}{c}{0.71}&
\multicolumn{1}{c}{0.97}&
\multicolumn{1}{c}{1.43}&
\multicolumn{1}{c}{1.46}&
\multicolumn{1}{c}{1.95}&
\multicolumn{1}{c|}{1.86}&
\multicolumn{1}{c}{\multirow{2}{*}{$10^5$}}\\
\multicolumn{1}{c|}{time 7L (sec)}&
\multicolumn{1}{c}{0.18}&
\multicolumn{1}{c}{0.18}&
\multicolumn{1}{c}{0.16}&
\multicolumn{1}{c}{0.17}&
\multicolumn{1}{c}{0.19}&
\multicolumn{1}{c|}{0.17}\\
\multicolumn{1}{c|}{speed-up factor}&
\multicolumn{1}{c}{4.02}&
\multicolumn{1}{c}{5.39}&
\multicolumn{1}{c}{8.72}&
\multicolumn{1}{c}{8.62}&
\multicolumn{1}{c}{10.91}&
\multicolumn{1}{c|}{10.84}\\\cline{1-8}
\multicolumn{1}{c|}{error ($\times\:10^{-3}$)}&
\multicolumn{1}{c}{1.2}&
\multicolumn{1}{c}{1.3}&
\multicolumn{1}{c}{1.2}&
\multicolumn{1}{c}{1.7}&
\multicolumn{1}{c}{1.4}&
\multicolumn{1}{c|}{1.2}&
\multicolumn{1}{c}{samples}\\
\multicolumn{1}{c|}{time MC (sec)}&
\multicolumn{1}{c}{10.07}&
\multicolumn{1}{c}{12.31}&
\multicolumn{1}{c}{19.35}&
\multicolumn{1}{c}{19.63}&
\multicolumn{1}{c}{24.85}&
\multicolumn{1}{c|}{24.64}&
\multicolumn{1}{c}{\multirow{2}{*}{$10^6$}}\\
\multicolumn{1}{c|}{time 7L (sec)}&
\multicolumn{1}{c}{1.99}&
\multicolumn{1}{c}{1.94}&
\multicolumn{1}{c}{1.86}&
\multicolumn{1}{c}{1.81}&
\multicolumn{1}{c}{1.95}&
\multicolumn{1}{c|}{1.96}\\
\multicolumn{1}{c|}{speed-up factor}&
\multicolumn{1}{c}{5.05}&
\multicolumn{1}{c}{6.33}&
\multicolumn{1}{c}{10.42}&
\multicolumn{1}{c}{10.84}&
\multicolumn{1}{c}{12.71}&
\multicolumn{1}{c|}{12.59}
\end{tabular}}
\end{center}
\end{table}\vspace{-0.7cm}

\subsubsection{Sampling from the SABR model}
\label{sssec: SABR}

\begin{figure}[b!]%
    \centering
    \subfloat[\centering]{{\includegraphics[width=7.cm]{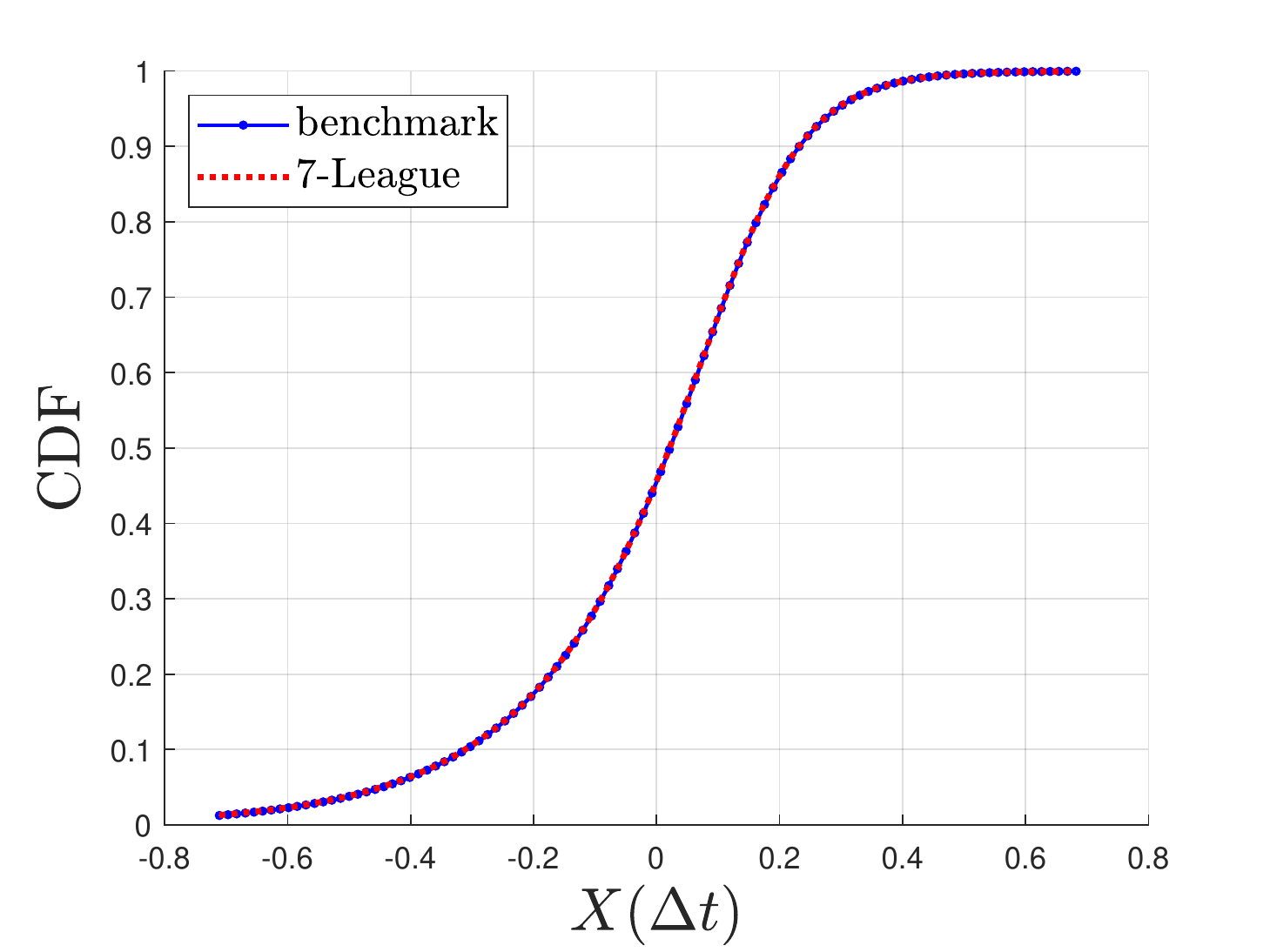} }}%
    ~
    \subfloat[\centering]{{\includegraphics[width=7.cm]{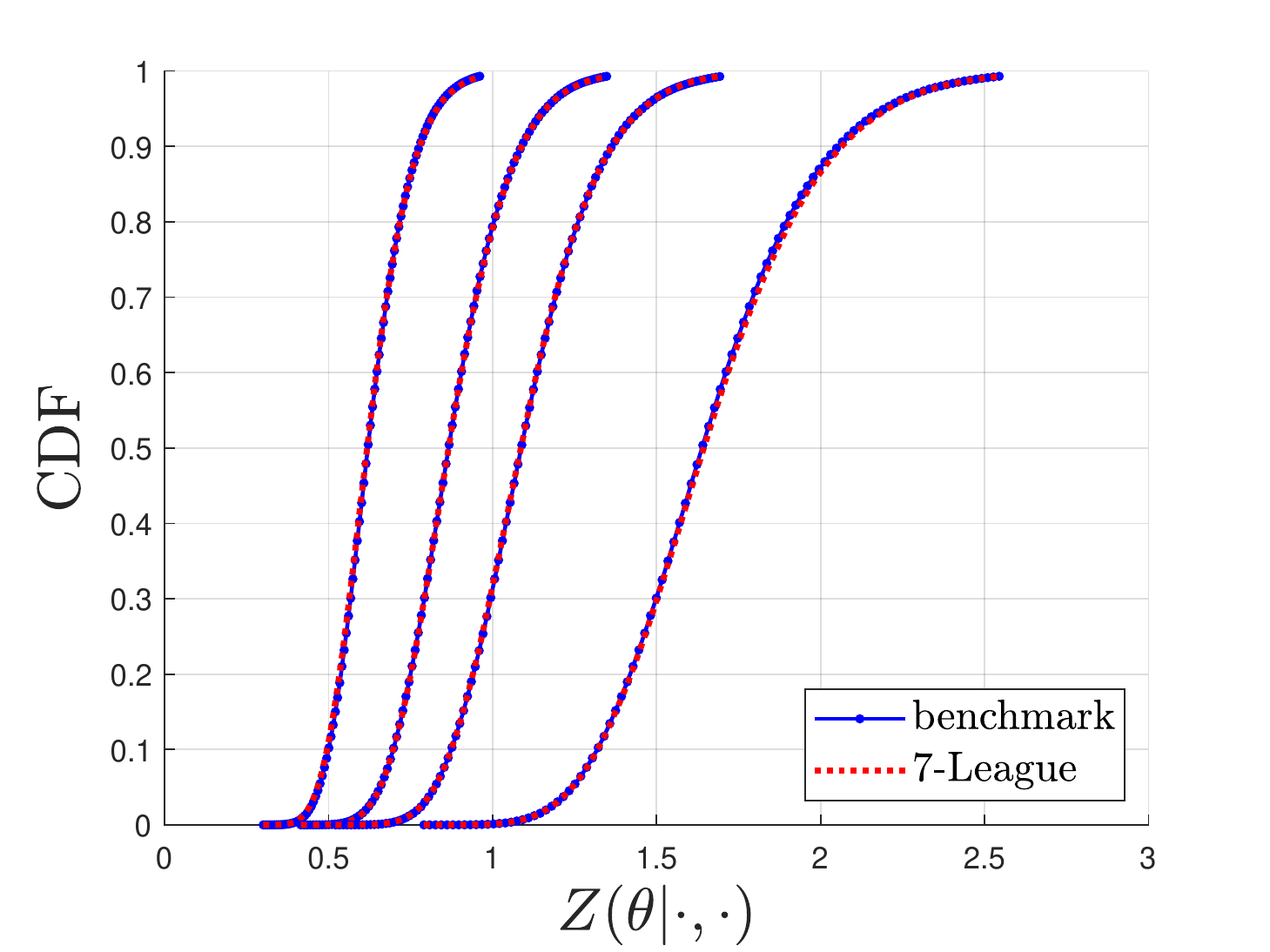} }}%
    \caption{Left: Empirical CDFs comparison for the log-price $X(\Dt)$ under the Heston model for parameters' Set V. Right: Empirical CDFs comparison for the conditional time-integrated squared volatility process, with boundary conditions $a=1$ and for $b$ respectively the quantiles 5\%, 35 \%, 65\% and 95\% (from the left).}%
    \label{fig: applications}%
\end{figure}

In this section we show how the methodology can be applied in the sampling under the SABR model assumptions. We only provide results about the sampling of the most tricky building-block needed in the \emph{almost exact} simulation presented in Section \ref{sssec: intro SABR}, which is the conditional time-integrated \emph{squared} volatility $\int_0^{\Dt}\sigma^2(t)\d t\big|\sigma(0), \sigma(\Dt)$.

Since the GBM in the SABR model is drift-less, to achieve a better approximation, the ANN is trained only on inputs $\theta=\{\alpha, \Dt\}$, with $\alpha$ diffusion parameter as in Section \ref{sssec: intro SABR}. The resulting empirical CDF for the time-integral of the squared volatility process is compared with a benchmark obtained via expensive MC simulation and is illustrated in Figure \ref{fig: applications}b. As we can see, even in this application, the benchmark CDF is almost indistinguishable from the one obtained through the methodology. For the full sampling from the SABR model (once the samples from the conditional time-integrated squared volatility process are available) see \cite{LeGrOo2, LeGrOo}.

\section{Conclusions}
\label{sec: conclusion}
In this article, we presented a robust data-driven procedure to sample from time-integrated stochastic bridges. The proposed methodology gives highly satisfactory results in terms of both accuracy and computational time, providing a powerful alternative to classical Monte Carlo schemes.
During the implementation, many hyper-parameters had to be tuned, and indeed better results can be achieved, in particular, if \emph{ad hoc} algorithms are developed for specific applications. In the article, we also provided insight about possible financial applications, but many others exist. 
Eventually, the idea underlying this methodology can be further explored to produce other alternatives to MC simulations for different classes of stochastic processes.

\nocite{*}
\bibliography{bibliography.bib}
\bibliographystyle{abbrv}

\end{document}